\documentclass[12pt]{article}
\pdfoutput=1
\usepackage{epsf,amsfonts,amssymb,epsfig,amsmath,mathtools,graphics,slashed,mathrsfs}
\usepackage{hyperref,hep,graphicx,subfig}
\usepackage{rotating}
\usepackage{dsfont}

\newcommand{\nn}{\notag \\}

\newcommand{\Om}{\Omega}

\newcommand{\K}{K}
\newcommand{\db}{\psi}

\def\sst#1{{\scriptscriptstyle #1}}
\def\0{{\sst{(0)}}}
\def\1{{\sst{(1)}}}
\def\2{{\sst{(2)}}}
\newcommand\3{{\sst{(3)}}}
\def\4{{\sst{(4)}}}
\def\5{{\sst{(5)}}}
\def\6{{\sst{(6)}}}
\def\7{{\sst{(7)}}}
\def\8{{\sst{(8)}}}

\def\ep{{\epsilon}}
\newcommand{\w}{\wedge}
\def\oneone{\rlap 1\mkern4mu{\rm l}}

\makeatletter

\@addtoreset{equation}{section}
\makeatother

\begin{document}

\begin{titlepage}

\vfill

\begin{flushright}
Imperial/TP/2019/JG/02\\
\end{flushright}

\vfill

\begin{center}
   \baselineskip=16pt
   {\Large\bf Consistent KK truncations for M5-branes\\ wrapped on Riemann surfaces}
  \vskip 1.5cm
K. C. Matthew Cheung$^1$, Jerome P. Gauntlett$^1$ and Christopher Rosen$^2$\\
     \vskip .6cm
             \begin{small}\vskip .6cm
      \textit{$^1$Blackett Laboratory, 
  Imperial College\\ London, SW7 2AZ, U.K.}
        \end{small}\\
             \begin{small}\vskip .6cm
      \textit{$^2$Departament de F\'\i sica Qu\`antica i Astrof\'\i sica and Institut de Ci\`encies del Cosmos (ICC),\\  Universitat de Barcelona, Mart\'\i\  i Franqu\`es 1, ES-08028, Barcelona, Spain.}
        \end{small}\\
                               \end{center}
\vfill

\begin{center}
\textbf{Abstract}
\end{center}
\begin{quote}
We construct a consistent Kaluza-Klein reduction of $D=11$ supergravity on $\Sigma_2\times S^4$, where 
$\Sigma_2=S^2,\mathbb{R}^2$ or $H^2$, or a quotient thereof, at the level of the bosonic fields. 
The result is a gauged $N=4$, $D=5$ supergravity theory coupled to three vector multiplets,
with the gauging lying in an 
$SO(2)\times SE(3)\subset SO(5,3)$ subgroup of the $SO(1,1)\times SO(5,3)$ global symmetry group of the ungauged theory.
For $\Sigma_2=H^2$, 
the $D=5$ theory has a maximally supersymmetric $AdS_5$ vacuum which uplifts to the known solution
of $D=11$ supergravity corresponding to M5-branes wrapping a Riemann surface 
with genus greater than one and dual to an $N=2$ SCFT in $d=4$.
For $\Sigma_2=S^2$, we find two $AdS_5$ solutions, one of which is new, and both of which are unstable.
There is an additional subtruncation to an $N=2$ gauged supergravity
coupled to two vector multiplets, with very special real manifold 
$SO(1,1)\times SO(1,1)$, and a single hypermultiplet, with quaternionic K\"ahler
manifold $SU(2,1)/S[U(2)\times U(1)]$ and gauging associated with an $SO(2)\times\mathbb{R}\subset SU(2,1)$ subgroup.

\end{quote}

\vfill

\end{titlepage}

\tableofcontents

\newcommand{\be}{\begin{equation}} \newcommand{\ee}{\end{equation}}
\newcommand{\bea}{\begin{eqnarray}} 
\newcommand{\eea}{\end{eqnarray}}

\setcounter{equation}{0}
\section{Introduction}

Consistent Kaluza-Klein truncations provide a powerful framework
for constructing solutions of $D=10$ and $D=11$ supergravity by solving the equations
of motion of a simpler supergravity theory in lower spacetime dimensions.
A particularly interesting setting is associated with supersymmetric $AdS_{d+1}\times M$
solutions since it allows one to study certain aspects of the dual SCFTs
from the gravitational side in a tractable way. Indeed, this framework has been
used to obtain many important results in holography
such as finding new fixed points, both with and without conformal invariance as well as constructing RG flows between them, constructing novel black holes dual to exotic strongly coupled states of matter and so on.

Given such an $AdS_{d+1}\times M$  solution, after carrying out a Kaluza-Klein reduction of the higher dimensional supergravity theory
on $M$, it is expected \cite{Gauntlett:2007ma}, and in several cases proven\footnote{There are some cases in which this has been proven in full generality, including the fermion fields, for example
\cite{Nastase:1999cb,Nastase:1999kf}. In other cases
it has been proven at the level of the bosonic fields.}, 
that it is always possible to truncate to a gauged supergravity in $d+1$ spacetime dimensions for which
the fields are dual to the superconformal current multiplet of the dual SCFT.
For example, associated with the maximally supersymmetric 
$AdS_7\times S^4$ and $AdS_4\times S^7$ solutions there are consistent KK truncations of $D=11$ supergravity on $S^4$ and $S^7$ down to maximally supersymmetric $SO(5)$ gauged supergravity in $D=7$ and
$SO(8)$ gauged supergravity in $D=4$, respectively \cite{Nastase:1999cb,Nastase:1999kf,deWit:1986oxb}. 
Similarly, associated with the 
maximally supersymmetric $AdS_5\times S^5$ solution there is a consistent truncation
of type IIB on $S^5$ down to maximally supersymmetric $SO(6)$ gauged supergravity in $D=5$
\cite{Cvetic:2000nc,Lee:2014mla,Baguet:2015sma}.

In this paper we present a new consistent KK truncation of $D=11$ supergravity on 
$\Sigma_2\times S^4$, where  $\Sigma_2=S^2,\mathbb{R}^2$ or $H^2$, or a quotient thereof, 
down to a half maximal gauged supergravity in $D=5$. One starting point for this result is the half maximal supersymmetric $AdS_5\times H^2/\Gamma\times S^4$ solution of 
\cite{Maldacena:2000mw}, where $H^2/\Gamma$ is a Riemann surface with genus greater than one, that are dual to $N=2$ SCFTs in
$d=4$. The $S^4$ factor is non-trivially fibred over the $H^2/\Gamma$ factor and correspondingly 
the solution describes the near horizon limit of M5-branes wrapping an $H^2/\Gamma$ factor, embedded inside a Calabi-Yau two-fold. An alternative point of view
is that the dual $N=2$, $d=4$ SCFTs are obtained by starting with the $N=(0,2)$, $d=6$ SCFT, dual to the $AdS_4\times S^7$ solution, compactifying on $H^2/\Gamma$ with a topological twist in order to preserve
$N=2$ supersymmetry in $d=4$, and then flowing to the IR.

Associated with this solution one should be able to compactify $D=11$ supergravity on 
$H^2/\Gamma\times S^4$ and truncate to the half-maximal $N=4$ Romans' $SU(2)\times U(1)$
gauged supergravity in $D=5$. In fact this result, at the level of the bosonic fields, was already obtained in \cite{Gauntlett:2007sm}. Here we will show that one can actually extend this truncation to an
$N=4$ gauged supergravity in $D=5$ coupled to three additional vector multiplets. 
We will carry out the KK truncation from $D=11$, first by reducing
on $S^4$ to maximal gauged supergravity in $D=7$ and then further reducing on the 
$H^2/\Gamma$ factor. The gauged supergravity that we construct contains the RG flow solution
described above, and first constructed in \cite{Maldacena:2000mw}, that is associated with the $N=(0,2)$ field theory in $d=6$ compactified on $H^2/\Gamma$ and flowing to an $N=2$ SCFT in $d=4$.

Furthermore, we show that one can also carry out a similar consistent KK truncation of $D=11$ supergravity on $\Sigma_2\times S^4$, where  $\Sigma_2=S^2,\mathbb{R}^2$ (or a quotient thereof). 
For these cases there is not a corresponding supersymmetric $AdS_5$ vacuum solution, 
which is certainly not a requisite for the existence of a consistent KK truncation, but the truncations still have a natural holographic interpretation. Indeed they incorporate the RG flows
associated with compactifying the $d=6$ $(0,2)$ SCFT on $S^2$ or $\mathbb{R}^2$, 
with, in the former case, a topological twist that preserves $N=2$ $d=4$ supersymmetry, and then
flowing to the IR \cite{Maldacena:2000mw}. Unlike the $H^2$ case, these theories do not flow to SCFTs in the IR.

We show that the consistent KK truncation of $D=11$ supergravity on 
$\Sigma_2\times S^4$ leads to an $N=4$, $D=5$ gauged supergravity with three vector multiplets and the gauging lying in an $SO(2)\times SE(3)\subset SO(5,3)$ subgroup of the $SO(1,1)\times SO(5,3)$ global symmetry group of the ungauged theory. One motivation for this work came from the possibility that the resulting $N=4$ gauged supergravity could have additional supersymmetric $AdS_5$ vacua and corresponding flows between them. Indeed, such scenarios in $N=4$
gauged supergravity were studied from a bottom up perspective in \cite{Bobev:2018sgr} and so it is
of considerable interest to investigate which of these scenarios can be realised in a top down setting.
Using the results of \cite{Bobev:2018sgr} we will show that the only maximally supersymmetric $AdS_5$ solution of the 
$N=4$, $D=5$ gauged supergravity theory that we obtain is the one that uplifts
to the $AdS_5\times H^2/\Gamma\times S^4$ solution of \cite{Maldacena:2000mw}.
We have also investigated the possibility of other $AdS_5$ solutions, supersymmetric or not. 
We find that the $N=4$, $D=5$ theory admits two non-supersymmetric
$AdS_5\times S^2\times S^4$ solutions, one of which was first found in
\cite{Gauntlett:2002rv}, while the other one is new. However, both of them have scalar modes that violate the BF bound and hence
are unstable. It is possible that there are additional $AdS_5$ solutions.

We also show that there are additional subtruncations of the $N=4$ gauged supergravity theory. 
When $\Sigma_2=H^2$ (and not $\Sigma_2=S^2,\mathbb{R}^2$) we can consistently truncate to Romans' gauged supergravity theory, as already 
mentioned above, and then further to minimal $D=5$ gauged supergravity. 
When $\Sigma_2=S^2,\mathbb{R}^2$ or $H^2$,
there is also a particularly interesting truncation
to an $N=2$, $D=5$ gauged supergravity theory coupled to two vector multiplets, with very special real manifold $SO(1,1)\times SO(1,1)$, and a single hypermultiplet, with quaternionic K\"ahler
manifold $SU(2,1)/S[U(2)\times U(1)]$, with the gauging associated with an $SO(2)\times\mathbb{R}\subset SU(2,1)$ subgroup.
A further truncation of this theory leads to a consistent truncation that was first constructed in \cite{Szepietowski:2012tb}.

The plan of the rest of the paper is as follows. In section \ref{maxsugra} we briefly recall
maximal $D=7$ gauged supergravity and how any bosonic solution can be uplifted to $D=11$.
In section \ref{sec3} we discuss the consistent KK truncation of maximal $D=7$ gauged
supergravity on $\Sigma_2$ and section \ref{sec4} shows, at the level of the bosonic fields, 
that the resulting $D=5$ theory is indeed an $N=4$ gauged supergravity theory. Section 
\ref{subtrunc} discusses some subtruncations and section \ref{secsols} discusses some
solutions, including the new and unstable $AdS_5\times  S^2\times S^4$ solution.
We conclude in section \ref{secfincom} and we have a few appendices which contain some useful results.

\section{Maximal $D=7$ gauged supergravity}\label{maxsugra}
Maximal gauged supergravity in $D=7$ \cite{Pernici:1984xx} has thirty two supercharges. 
The bosonic fields consist of a metric, $SO(5)$
Yang-Mills one-form potentials $A^{ij}$, $i,j=1,\dots 5$ transforming in the ${\bf 10}$ of $SO(5)$, three-forms $S_{(3)}^i$ transforming in the  ${\bf 5}$,  and fourteen scalar fields, given by the symmetric unimodular matrix
$T_{ij}$, which parametrise the coset $SL(5,\mathbb{R})/SO(5)$.
The seven-form Lagrangian for the bosonic fields is given by
\begin{align}
{\cal L} &= R\, {*\oneone} -
\tfrac{1}{4} T^{-1}_{ij}\, {*D T_{jk}}\wedge
T^{-1}_{k l}\, D T_{l i}
-\tfrac{1}{4}\, T^{-1}_{ik}\, T^{-1}_{jl}\, {* F_\2^{ij}}\wedge F_\2^{kl}
-\tfrac{1}{2} T_{ij}\, {*S_\3^i}\wedge S_\3^j \nn
&+ \tfrac{1}{2g} S_\3^i\wedge DS_\3^i -
\tfrac{1}{8g}  \ep_{i j_1\cdots j_4}\, S_\3^i\wedge F_\2^{j_1 j_2}\wedge
F_\2^{j_3 j_4} +
\tfrac{1}{g} \Omega_\7 - V\, {*\oneone}\,,\label{d7lag}
\end{align}
with
\begin{align}
DT_{ij}& \equiv dT_{ij} + g A_\1^{ik}\, T_{kj} + g A_\1^{jk}\, T_{ik}\,,\nn
D S_\3^i &\equiv dS_\3^i + g\, A_\1^{ij}\wedge S_\3^j\,,\nn
F_\2^{ij} &\equiv dA_\1^{ij} + g A_\1^{ik}\wedge A_\1^{kj}\,,
\end{align}
where $g$ is a coupling constant.
The potential $V$ is given by
\begin{align}
V = \tfrac{1}{2}  g^2 \Big(2 T_{ij}\, T_{ij} - (T_{ii})^2 \Big)\,,
\end{align}
and $\Omega_\7$ is a Chern-Simons type of term built from the
Yang-Mills fields, which has the property that its variation with
respect to $A_\1^{ij}$ gives
\begin{align}
\delta \Omega_\7 =
\tfrac{3}{4} \delta_{i_1 i_2 k l}^{j_1 j_2 j_3 j_4}\, F_\2^{i_1 i_2}\wedge
F_\2^{j_1 j_2}\wedge  F_\2^{j_3 j_4}\wedge \delta A_\1^{k l}\,.
\end{align}
An explicit expression can be found in \cite{Pernici:1984xx}.

Any solution to the associated $D=7$ equations of motion, which are given in
appendix A, gives rise to a solution of $D=11$ supergravity
\cite{Nastase:1999cb,Nastase:1999kf}. Using the notation of
\cite{Cvetic:2000ah}, the $D=11$ metric and four-form field strength are given by
\begin{align}
d s_{11}^2 &= \Delta^{1/3}\, ds_{7}^2 + \frac{1}{g^2}\Delta^{-2/3}\,
T^{-1}_{ij}\, D\mu^i\, D\mu^j\,,\label{metel}
\end{align}
\begin{align}
G_\4 &= \frac{\Delta^{-2}}{g^34!}\, \ep_{i_1\cdots i_5}\, \Big[
-  U\,  \mu^{i_1} D\mu^{i_2}\wedge D\mu^{i_3}\wedge D\mu^{i_4}\wedge
D\mu^{i_5}\nn
& + 4 \, T^{i_1 m}\, DT^{i_2 n}\, \mu^m\, \mu^n\,
D\mu^{i_3}
\wedge D\mu^{i_4} \wedge D\mu^{i_5}+ 6g \Delta F_\2^{i_1 i_2} \wedge
D\mu^{i_3}\wedge D\mu^{i_4}\, T^{i_5 j}\, \mu^j \Big]\nn
&- T_{ij}\,
{\ast S_\3^i}\, \mu^j + \frac{1}{g}\, S_\3^i \wedge D\mu^i\,,\label{4form}
\end{align}
where $\mu^i= 1, \ldots, 5$ are constrained coordinates on $S^4$ satisfying $\mu^i \mu^i =1$, and
\bea
U \equiv 2 T_{ij}\, T_{jk}\, \mu^i\, \mu^k - \Delta\, T_{ii}\,, \qquad
\Delta \equiv T_{ij}\, \mu^i\, \mu^j\,,\qquad
D\mu^i \equiv d\mu^i + g A_\1^{ij}\, \mu^j \,.
\eea

The $AdS_7$ vacuum solution of $D=7$ supergravity with $A^{ij}_\1=S^i_\3=0$ and $T_{ij}=\delta_{ij}$, preserves all of the supersymmetry and uplifts to the maximally supersymmetric $AdS_7\times S^4$ solution, arising as the near horizon limit of a stack of M5-branes. In \cite{Maldacena:2000mw}
two different supersymmetric $AdS_5\times H^2$ solutions were found 
which uplift to $AdS_5\times H^2\times S^4$ solutions, with a warped product metric and
the $S^4$ non-trivially fibred over the $H^2$ factor. The fibration structure differs in the two solutions of \cite{Maldacena:2000mw}
and they either preserve 16 or 8 supercharges. In each case the
$H^2$ factor can be replaced with an arbitrary quotient $H^2/\Gamma$, while preserving supersymmetry, and we are particularly
interested in the case when $H^2/\Gamma$ is a compact Riemann surface with genus greater than one.
The solutions are dual to $N=2$ or $N=1$ superconformal field theories in four spacetime dimensions, respectively,
that arise on the non-compact part of M5-branes wrapping such a Riemann surface that is holomorphically embedded
either in a Calabi-Yau two-fold or three-fold, respectively. In this paper, it is the solution preserving 16 supercharges, which is recorded in section \ref{susvacsec}, that is of relevance. In particular, we will use the fibration structure
of this solution to construct a new consistent KK truncation of maximal $D=7$ gauged supergravity reduced on $H^2$ as well as on $S^2$ and $\mathbb{R}^2$. We note that it is only the $H^2$ case that the $D=5$ theory 
has a maximally supersymmetric $AdS_5$ vacuum solution. 
For the $S^2$ case
there is a non-supersymmetric $AdS_5$ solution found \cite{Gauntlett:2002rv} as well as an additional
new solution that we discuss in section \ref{nonsusy}.

\section{Consistent KK truncation on $S^2, \mathbb{R}^2$ or $H^2$}\label{sec3}

We now construct the consistent KK ansatz for the reduction of maximal $D=7$ gauged supergravity on 
$\Sigma_2=S^2, \mathbb{R}^2$ or $H^2$, or a quotient thereof.  

\subsection{The consistent truncation}
The ansatz for the $D=7$ metric is given by
\begin{align}\label{metans}
ds^2_7=e^{-4\phi}ds^2_5+e^{6\phi} ds^2(\Sigma_2)\,,
\end{align}
where $\phi$ is a scalar field defined on the five-dimensional spacetime. We introduce an orthonormal frame
for the two-dimensional metric and write 
$ds^2(\Sigma_2)=\bar e^a\bar e^a$ and $d\bar e^a+\bar\omega^a{}_b\wedge \bar e^b=0$, with $a,b=1,2$.
We normalise this metric so that 
$R^{(2)}_{ab}=lg^2\delta_{ab}$, with $l=1,0,-1$ for $\Sigma_2=S^2, \mathbb{R}^2$ or $H^2$, respectively.
We also write $\mathrm{vol}(\Sigma_2)=\bar e^1\wedge \bar e^2$.

We decompose the $D=7$ $SO(5)$ gauge fields via
$SO(5)\to SO(2)\times SO(3)$ and write
\begin{align}\label{gfans}
A^{ab}_{(1)}&=\tfrac{1}{g}\bar{\omega}^{ab}+\epsilon^{ab}A_{(1)}\,,\nn
A^{a\alpha}_{(1)}&=-A^{\alpha a}_{(1)}=\db^{1\alpha} \bar{e}^a-\db^{2\alpha}\epsilon^{ab} \bar{e}^b\,,\nn
A^{\alpha\beta}_{(1)}&=A^{\alpha \beta}_{(1)}\,,
\end{align}
with $a,b=1,2$ and $\alpha,\beta=3,4,5$. Crucially, this ansatz is anchored by the spin connection, $\bar\omega^{ab}$,
of $\Sigma_2$ in the expression for $A^{ab}$ which, in particular, allows one to study M5-branes wrapping Riemann surfaces
with a ``topological twist" so that $N=2$, $d=4$ supersymmetry is preserved on the non-compact part of the M5-brane worldvolume.
The ansatz \eqref{gfans} introduces an $SO(2)$ one-form $A_{(1)}$,  $SO(3)$ one-forms
$A_{(1)}^{\alpha\beta}$ transforming in the $({\bf 1},{\bf 3})$ of  $SO(2)\times SO(3)$, and six scalars $\db^{a\alpha}\equiv (\db^{1\alpha},\db^{2\alpha})$, transforming as $({\bf{2}},{\bf 3})$, all
defined on the five-dimensional spacetime.
For the scalar fields we take
\begin{align}
T^{ab}=e^{-6\lambda}\delta^{ab}\,,\qquad
T^{a\alpha}=0\,,\qquad
T^{\alpha\beta}=e^{4\lambda} \mathcal{T}^{\alpha\beta}\,,
\end{align}
which introduces a $D=5$ scalar $\lambda$ as well as another five scalars in the symmetric, unimodular
matrix $\mathcal{T}^{\alpha\beta}$ which parametrise the coset $SL(3)/SO(3)$.
For the $D=7$ three-form we take
\begin{align}\label{3fans}
S^a_{(3)}&=\K^1_{(2)}\wedge\bar{e}^a-\epsilon^{ab}\K^2_{(2)}\wedge \bar{e}^b\,,\nn
S^\alpha_{(3)}&=h^{\alpha}_{(3)}+\chi^{\alpha}_{(1)}\wedge \mathrm{vol}(\Sigma_2)\,,
\end{align}
giving rise in $D=5$ to an $SO(2)$ doublet of two-forms $\K^a_{(2)}\equiv (\K^1_{(2)},\K^2_{(2)})$ transforming as $({\bf 2},{\bf 1})$, as well as $({\bf 1},{\bf 3})$ three-forms $h_{(3)}^\alpha$ and $({\bf 1},{\bf 3})$
one-forms $\chi_{(1)}^\alpha$. Finally, for later convenience, for the $D=5$ fields instead of taking the indices
$\alpha,\beta,\gamma,\dots \in \{3,4,5\}$ we will take 
\begin{align}\label{relabalphind}
\alpha,\beta,\gamma,\dots \in \{1,2,3\}\,.
\end{align}

We can substitute this ansatz into the $D=7$ equations of motion. After some
long calculation we can show that they are equivalent to a set of unconstrained equations of motion for the $D=5$ fields, which shows that
the truncation is consistent. Some details of this calculation is presented in appendix \ref{seca1} and the final $D=5$ equations of motion
are recorded in \eqref{redeq1}-\eqref{redeq2} and \eqref{redeq3}-\eqref{redeq10}.
Moreover, these $D=5$ equations of motion can be derived from a five-form  Lagrangian given by
\begin{equation}
\mathcal{L}={R}\mathrm{vol}_{5}+\mathcal{L}^{kin}+\mathcal{L}^{pot}+\mathcal{L}^{top}\,,
\end{equation}
where $R$ is the Ricci scalar of the $D=5$ metric and the remaining
kinetic energy terms are 
\begin{align}\label{ellkin}
\mathcal{L}^{kin}=&-30{\ast{} d\phi}\wedge{d\phi}-30{\ast{} d\lambda}\wedge{d\lambda}
-\tfrac{1}{4}\mathcal{T}^{-1}_{\alpha\beta}\mathcal{T}^{-1}_{\gamma\rho}{\ast  D\mathcal{T}_{\beta\gamma}}\wedge {D \mathcal{T}_{\rho\alpha}}\nn
&-\tfrac{1}{2}e^{12\lambda+4\phi}{\ast  F_{(2)}}\wedge F_{(2)}-e^{-6\lambda-2\phi}{\ast  \K^a_{(2)}}\wedge \K^a_{(2)}
\nn
&-\tfrac{1}{4}e^{-8\lambda+4\phi}\mathcal{T}^{-1}_{\alpha\beta}\mathcal{T}^{-1}_{\gamma\rho} {\ast  {F}^{\alpha\gamma}_{(2)}}\wedge {F}^{\beta\rho}_{(2)}
-e^{2\lambda-6\phi}\mathcal{T}^{-1}_{\alpha\beta}{\ast  D\db^{a\alpha}}\wedge D\db^{a\beta}\nn
&-\tfrac{1}{2}e^{4\lambda-12\phi}\mathcal{T}_{\alpha\beta}{\ast  \chi^\alpha_{(1)}}\wedge \chi^{\beta}_{(1)}-\tfrac{1}{2}e^{4\lambda+8\phi}\mathcal{T}_{\alpha\beta}{\ast  h^\alpha_{(3)}}\wedge h^{\beta}_{(3)}\,.
\end{align}
The potential terms are
\begin{align}\label{ellpot}
\mathcal{L}^{pot}=&\,{g}^2\Big\{-\tfrac{1}{2}e^{12\lambda-16\phi}(l-\db^2)^2
-e^{-8\lambda-16\phi}
\epsilon^{ab}\epsilon^{cd}(\db^{a}\mathcal{T}^{-1}\db^{c})
(\db^{b}\mathcal{T}^{-1}\db^{d})
\nn
&+e^{-10\phi}\left(2(l+\db^2)
-e^{10\lambda}(\db\mathcal{T}\db)
-e^{-10\lambda}(\db\mathcal{T}^{-1}\db)\right)\nn
&
+\tfrac{1}{2}e^{-4\phi}\left(e^{8\lambda}(\mathrm{Tr}\mathcal{T})^2-2e^{8\lambda}\mathrm{Tr}(\mathcal{T}^2)+4e^{-2\lambda}\mathrm{Tr}\mathcal{T}\right)
\Big\}\mathrm{vol}_{5}\,,
\end{align}
where $\psi^2\equiv \psi^{a\alpha}\psi^{a\alpha}$
and the topological term, independent of the $D=5$ metric, is given by
\begin{align}\label{elltop}
\mathcal{L}^{top}=&\,\tfrac{1}{g}
\epsilon^{ab}\K^a_{(2)}\wedge\left(D\K^b_{(2)}-g \db^{b\alpha}h^{\alpha}_{(3)}\right)
+\tfrac{1}{g}\epsilon_{\alpha\beta\gamma}\K^a_{(2)}\wedge D\db^{a\gamma}\wedge{F}^{\alpha\beta}_{(2)}\nn
&+\tfrac{1}{2g}h^{\alpha}_{(3)}\wedge\left(D\chi^{\alpha}_{(1)}
+2g\epsilon^{ab}\db^{a\alpha} \K^b_{(2)}\right)+\tfrac{1}{2g}\chi^{\alpha}_{(1)}\wedge Dh^{\alpha}_{(3)}\nn
&-\tfrac{1}{2}\epsilon_{\alpha\beta\gamma}(l-\db^2)h^{\alpha}_{(3)}\wedge {F}^{\beta\gamma}_{(2)}
-\epsilon_{\alpha\beta\gamma}(\epsilon^{ab}\db^{a\beta}\db^{b\gamma})h^{\alpha}_{(3)}\wedge F_{(2)}
\nn
&-\tfrac{1}{2g}\epsilon_{\alpha\beta\gamma}\chi^{\alpha}_{(1)}\wedge {F}^{\beta\gamma}_{(2)}\wedge F_{(2)}
-\tfrac{1}{g}\epsilon_{\alpha\beta\gamma}h^{\alpha}_{(3)}\wedge D\db^{a\beta}\wedge D\db^{a\gamma}\nn
&+\tfrac{1}{g}(\db^{a\alpha} D\db^{a\beta})\wedge {F}^{\alpha\beta}_{(2)}\wedge F_{(2)}
+\tfrac{1}{2g}(\epsilon^{ab}\db^{a\gamma} D\db^{b\gamma})\wedge {F}^{\alpha\beta}_{(2)}\wedge{F}^{\alpha\beta}_{(2)}
\nn
&+\tfrac{1}{2}l \,{F}^{\alpha\beta}_{(2)}\wedge {F}^{\alpha\beta}_{(2)}\wedge A_{(1)}
-\tfrac{1}{g}(\epsilon^{ab}\db^{a\alpha} D\db^{b\beta})\wedge{F}^{\alpha\gamma}_{(2)}\wedge{F}^{\beta\gamma}_{(2)}\,.
\end{align}
In these expressions we have used the following definitions of 
field strengths and covariant derivatives:
\begin{align}
F_{(2)}&\equiv dA_{(1)}\,,\qquad
{F}^{\alpha\beta}_{(2)}\equiv dA^{\alpha\beta}_{(1)}+gA^{\alpha\gamma}_{(1)}\wedge A^{\gamma\beta}_{(1)}\,,\nn
D\db^{a\alpha}&\equiv d\db^{a\alpha}+gA^{\alpha\beta}_{(1)}\db^{a\beta}+gA_{(1)}\epsilon^{ab}\db^{b\alpha}\,,
\qquad
D\mathcal{T}_{\alpha\beta}\equiv d\mathcal{T}_{\alpha\beta}+gA^{\alpha\gamma}_{(1)}\mathcal{T}_{\gamma\beta}+gA^{\beta\gamma}_{(1)}\mathcal{T}_{\alpha\gamma}\,,\nn
D\K^a_{(2)}&\equiv d\K^a_{(2)}+g\epsilon^{ab}A_{(1)}\wedge \K^b_{(2)}\,,\nn
Dh^\alpha_{(3)}&\equiv dh^\alpha_{(3)}+gA^{\alpha\beta}_{(1)}\wedge h^\beta_{(3)}\,,\qquad
D\chi^{\alpha}_{(1)}\equiv d\chi^\alpha_{(1)}+gA^{\alpha\beta}_{(1)}\wedge \chi^{\beta}_{(1)}\,.
\end{align}

\subsection{Field redefinitions}\label{fieldredef}
In order to make contact with half maximal $N=4$, $D=5$ supergravity in the next section, it is necessary to make a 
number of field redefinitions. We first define 
\begin{align}
A^{\alpha\beta}_{(1)}=\epsilon_{\alpha\beta\gamma}A^{\gamma}_{(1)}\,,
\end{align}
with the field strength for $A^{\alpha}_{(1)}$ given by
${F}^{\alpha}_{(2)}\equiv dA^{\alpha}_{(1)}-\tfrac{1}{2}g\epsilon_{\alpha\beta\gamma}
A^{\beta}_{(1)}\wedge A^{\gamma}_{(1)}$.
We next replace the one-form $\chi^\alpha_{(1)}$ with 
a one-form $\mathscr{A}_{(1)}^\alpha$ and three Stueckelberg scalar fields
$\xi^\alpha$, both transforming under $SO(3)$ in the triplet representation,  
via
\begin{align}\label{chiredef}
\chi^\alpha_{(1)}=&\,D\xi^\alpha+g\mathscr{A}_{(1)}^\alpha+\epsilon_{\alpha\beta\gamma}\psi^{a\beta}D\psi^{a\gamma}\,,
\end{align}
with $D\xi^\alpha\equiv d\xi^\alpha-g\epsilon_{\alpha\beta\gamma}A^\beta_{(1)}\xi^\gamma$. Furthermore, the field redefinition introduces a new gauge invariance, with non-compact group, in which 
$\delta\xi^\alpha=\Lambda^\alpha(x)$, $\delta \mathscr{A}_{(1)}^\alpha=-g^{-1}D\Lambda^\alpha$, leaving
$\chi^\alpha_{(1)}$ invariant. This could be
used to eliminate the scalars $\xi^\alpha$ if desired. If we substitute this into
the equation of motion \eqref{redeq2} we deduce that 
\begin{align}
{\ast h^\alpha_{(3)}}=&\,e^{-4\lambda-8\phi}\mathcal{T}^{-1}_{\alpha\beta}\left(G^\beta_{(2)}+2\epsilon_{ab}\psi^{a\beta}K^b_{(2)}+\left(\epsilon_{\beta\gamma\rho}\xi^\gamma+\psi^{a\beta}\psi^{a\rho}\right){F}^\rho_{(2)}\right)\,,
\label{Eq. h3 redefinition in Vector and Topological stuff}
\end{align}
where we have defined the two-form
\begin{align}\label{xi_G_definition}
G^\alpha_{(2)}&\equiv D\mathscr{A}_{(1)}^\alpha - l{F}^\alpha_{(2)}\,,
\end{align}
with $D\mathscr{A}_{(1)}^\alpha\equiv d\mathscr{A}_{(1)}^\alpha-g\epsilon_{\alpha\beta\gamma}A^\beta_{(1)}\wedge \mathscr{A}_{(1)}^\gamma$. Notice that this expression for $h^{\alpha}_{(3)}$ is invariant under the new non-compact gauging
just mentioned. In carrying out the identification with the fields of gauged $N=4$ supergravity in the next
section, it is helpful to notice that we can also write
\begin{align}\label{xi_G_definitionagain}
G^\alpha_{(2)}=d(\mathscr{A}_{(1)}^\alpha-l{A}_{(1)}^\alpha)
-g\epsilon_{\alpha\beta\gamma}A^\beta\wedge (\mathscr{A}_{(1)}^\gamma-l{A}_{(1)}^\gamma)
-\frac{gl}{2}\epsilon_{\alpha\beta\gamma}A^\beta_{(1)}\wedge A^\gamma_{(1)}\,.
\end{align}
We also redefine the two-forms via
\begin{align}
K^a_{(2)}=&-\frac{1}{\sqrt{2}}\epsilon_{ab}L^b_{(2)}+\epsilon_{ab}\psi^{b\alpha}{F}^\alpha_{(2)}\,,
\end{align}
and finally exchange the two scalars $\phi,\lambda$ 
for two scalars $\varphi_3,\Sigma$ via
\begin{equation}\label{scalidents}
\varphi_3 = 3\phi-\lambda\,,\qquad \Sigma = e^{-(\phi+3\lambda)}\,.
\end{equation}

With these field redefinitions we find that the equations of motion given in
\eqref{redeq1}-\eqref{redeq2} and \eqref{redeq3}-\eqref{redeq10}
can be obtained from a Lagrangian of the form
\begin{align}\label{lagfred1}
\mathcal{L}=&\,{R}\mathrm{vol}_{5}+\mathcal{L}^S+\mathcal{L}^{pot}+\mathcal{L}^V+\mathcal{L}^T\,,
\end{align}
with the scalar kinetic terms given by
\begin{align}\label{lscalred}
\mathcal{L}^S=
&-3\Sigma^{-2}{\ast d\Sigma}\wedge d\Sigma
-3{\ast d\varphi_3}\wedge d\varphi_3
-\tfrac{1}{4}\mathcal{T}^{-1}_{\alpha\beta}\mathcal{T}^{-1}_{\gamma\rho}{\ast  D\mathcal{T}_{\beta\gamma}}\wedge {D \mathcal{T}_{\rho\alpha}}
\nn
&
-e^{-2\varphi_3}\mathcal{T}^{-1}_{\alpha\beta}{\ast  D\db^{a\alpha}}\wedge D\db^{a\beta}-\tfrac{1}{2}e^{-4\varphi_3}\mathcal{T}_{\alpha\beta}{\ast \chi^\alpha_{(1)}}\wedge 
 \chi^\beta_{(1)}\,,
\end{align}
after substituting for $\chi^\alpha_{(1)}$ using \eqref{chiredef}. The potential terms for the scalars 
are as in \eqref{ellpot} and can be written in terms of the new fields as 
\begin{align}\label{ellpotnf}
\mathcal{L}^{pot}=&\,{g}^2\Big\{
\Sigma^4\left(-e^{-4\varphi_3}
\epsilon^{ab}\epsilon^{cd}(\db^{a}\mathcal{T}^{-1}\db^{c})
(\db^{b}\mathcal{T}^{-1}\db^{d})
-e^{-2\varphi_3}(\db\mathcal{T}^{-1}\db)
\right)\nn
&+\Sigma^{-2}\left(
-\tfrac{1}{2}e^{-6\varphi_3}(l-\db^2)^2
-e^{-4\varphi_3}(\db\mathcal{T}\db)
+e^{-2\varphi_3}[ \tfrac{1}{2}(\mathrm{Tr}\mathcal{T})^2-\mathrm{Tr}(\mathcal{T}^2)  ]
\right)\nn
&+2\Sigma\left(
e^{-3\varphi_3}(l+\db^2)
+e^{-\varphi_3}\mathrm{Tr}\mathcal{T}
\right)
\Big\}\mathrm{vol}_{5}\,,
\end{align}
and we note, in particular, that the scalar potential is independent of the scalars $\xi^\alpha$.
The kinetic terms for the vectors 
are given by
\begin{align}\label{Eq. new vector ke completr}
&\mathcal{L}^{V}
=-\tfrac{1}{2}\Sigma^{-4}{\ast F_{(2)}}\wedge F_{(2)}\nn
&-\tfrac{1}{2}\Sigma^2\Big\{e^{-2\varphi _3}\mathcal{T}^{-1}_{\alpha\beta}{\ast G^\alpha_{(2)}}\wedge G^\beta_{(2)}+2\sqrt{2}e^{-2\varphi _3}\mathcal{T}^{-1}_{\alpha\beta}\psi^{a\beta}{\ast G^\alpha_{(2)}}\wedge {L}^a_{(2)}\nn
&
\qquad
-2e^{-2\varphi _3}\mathcal{T}^{-1}_{\alpha\beta}\left(\epsilon_{\beta\gamma\rho}\xi^\rho+\psi^{a\beta}\psi^{a\gamma}\right){\ast G^\alpha_{(2)}}\wedge {F}^\gamma_{(2)}\nn
&\qquad
-2\sqrt{2}\left(e^{-2\varphi _3}\psi^{a\beta}\mathcal{T}^{-1}_{\beta\gamma}\left(\epsilon_{\gamma\alpha\rho}\xi^\rho+\psi^{a\gamma}\psi^{a\alpha}\right)+\psi^{a\alpha}\right){\ast {L}^a_{(2)}}\wedge{F}^\alpha_{(2)}\nn
&
\qquad+\left(e^{2\varphi_3}\mathcal{T}_{\alpha\beta}+2\psi^{a\alpha}\psi^{a\beta}+e^{-2\varphi_3}\left(\epsilon_{\gamma\alpha\eta}\xi^\eta+\psi^{a\gamma}\psi^{a\alpha}\right)\mathcal{T}^{-1}_{\gamma\rho}\left(\epsilon_{\rho\beta\tau}\xi^\tau+\psi^{b\rho}\psi^{b\beta}\right)\right){\ast {F}^\alpha_{(2)}}\wedge{F}^\beta_{(2)}
\nn
&
\qquad
+\left(2e^{-2\varphi _3}\psi^{a\alpha}\mathcal{T}^{-1}_{\alpha\beta}\psi^{b\beta}+\delta_{ab}\right){\ast {L}^a_{(2)}}\wedge{L}^b_{(2)}
\Big\}\,.
\end{align}
Finally the remaining topological terms are given by the remarkably simple expression
\begin{equation}\label{eq:new_top_after_field_redef}
\mathcal{L}^{T}=\tfrac{1}{2g}\epsilon_{ab}{L}^a_{(2)}\wedge D{L}^b_{(2)}-G^\alpha_{(2)}\wedge{F}^\alpha_{(2)}\wedge A_{(1)}\,.
\end{equation}

\section{Supersymmetry}\label{sec4}
We now show that the reduced $D=5$ theory obtained in the previous section
is precisely the bosonic sector of an $N=4$ gauged supergravity in $D=5$, with sixteen supercharges, coupled to three vector multiplets.

\subsection{$N=4$ gauged supergravity}\label{sec4pt1}
In this subsection we first summarise the general structure of $N=4$ gauged supergravity in $D=5$, coupled to $n=3$ vector multiplets, mostly following the conventions and presentation of \cite{Schon:2006kz} (which generalised \cite{DallAgata:2001wgl}). 

We begin by recalling that the ungauged theory \cite{Awada:1985ep}
has a global symmetry group given by 
$SO(1,1)\times SO(5,n=3)$. The bosonic field content consists of a metric, 
$6+n=9$ Abelian vector fields and $1+5n=16$ scalar fields. The nine vector fields can be written as
$\mathcal{A}^0_{(1)}$ and $\mathcal{A}^M_{(1)}$, with $M=1,\dots,8$, which transform as a 
scalar and vector with respect to $SO(5,3)$, respectively. The scalar manifold is given by $SO(1,1)\times SO(5,3)/(SO(5)\times SO(3))$, with the
$SO(1,1)$ part described by a real scalar field $\Sigma$, while we parametrise 
the coset $SO(5,3)/(SO(5)\times SO(3))$ by the $8\times 8$ matrix $\mathcal{V}^A{}_M$. 
The matrix $\mathcal{V}^A{}_M$ is an element of $SO(5,3)$ satisfying
\begin{align}
\mathcal{V}^T\eta \mathcal{V} =\eta\,,
\end{align}
where $\eta$ is the invariant metric tensor of $SO(5,3)$. Global $SO(5,3)$ transformations are
taken to act on the right, while local $SO(5)\times SO(3)$ transformations act on the left via
\begin{align}
\mathcal{V}\to h(x)\mathcal{V}g \,,\quad\qquad g\in SO(5,3)\,,\quad h\in SO(5)\times SO(3)\,.
\end{align}
The coset can also be parametrised by a symmetric positive definite matrix $\mathcal{M}_{MN}$ defined by
\begin{align}\label{emmmat}
\mathcal{M}_{MN}=(\mathcal{V}^T \mathcal{V})_{MN}\,,
\end{align}
with $\mathcal{M}_{MN}$ an element of $SO(5,3)$. We can raise indices using $\eta$ and in particular
the inverse, which we denote by $\mathcal{M}^{MN}$, is given by 
\begin{equation}
\mathcal{M}^{MN}\equiv \eta^{MP}\eta^{NQ}\mathcal{M}_{PQ}=\left(\mathcal{M}^{-1}\right)^{MN}\,.
\end{equation}

We will work in a basis in which $\eta$ is not diagonal, but instead given by
\begin{equation}\label{etadef}
\eta = 
\begin{pmatrix}
0 & 0 & \mathds{1}_3\\
0 & -\mathds{1}_2 & 0\\
\mathds{1}_3 & 0 & 0
\end{pmatrix}\,.
\end{equation}
In order to work in a basis in which $\eta$ is diagonal with the first five entries $-1$ and the last three entries $+1$,
as in \cite{Schon:2006kz}, we can employ a similarity transformation using the matrix
\begin{equation}\label{usimmat}
\mathcal{U} =
\begin{pmatrix}
-U & 0 & U\\
0 & \mathds{1}_2 & 0\\
U & 0 & U
\end{pmatrix}\,,
\qquad \text{with}\qquad
U=\frac{1}{\sqrt{2}}\begin{pmatrix}
0 & 0 & 1\\
0 & 1 & 0\\
1& 0 & 0
\end{pmatrix}\,,
\end{equation}
which satisfies $\mathcal{U} = \mathcal{U}^T =\mathcal{U}^{-1}$ and $\det\mathcal{U}=1$. In the expression for the scalar potential in the
gauged theory, given below, we will also need the following antisymmetric tensor
\begin{align}\label{mfiveddefu}
\mathcal{M}_{M_1\dots M_5}\equiv \epsilon_{m_1\dots m_5}(\mathcal{U}\cdot\mathcal{V})^{m_1}{}_{M_1}\dots
(\mathcal{U}\cdot\mathcal{V})^{m_5}{}_{M_5}\,,
\end{align}
with the indices $m_1,\dots, m_5$ running from 1 to 5.

The general $N=4$, $D=5$ gauged theory \cite{Schon:2006kz}
is specified by a set of embedding tensors $f_{MNP}=f_{[MNP]}$, $\xi_{MN}=\xi_{[MN]}$ and $\xi_M$. These specify both the gauge group in $SO(1,1)\times SO(5,3)$
as well assigning specific vector fields to the generators of the gauge group. The covariant derivative is given by\footnote{\label{foot2}Here the terms involving the generators differ by a factor two with the analogous
expression in \cite{Schon:2006kz}. However, the explicit expression for the generators that we use 
in \eqref{explicgen} below, also differ by a factor of two implying that our
covariant derivative is the same as \cite{Schon:2006kz}.}
\begin{align}\label{eq_cov_derivative}
D_\mu&=\nabla_\mu-\tfrac{1}{2}g\left(\mathcal{A}^M_{(1)\mu}f_{M}^{\phantom{M}NP}t_{NP}+\mathcal{A}^0_{(1)\mu}\xi^{NP}t_{NP}
+\mathcal{A}^M_{(1)\mu}\xi^Nt_{MN}+\mathcal{A}^M_{(1)\mu}\xi_{M}t_{0}\right)\,,
\end{align}
where $t_{MN}=t_{[MN]}$ are the generators for $SO(5,3)$, $t_0$ is the generator for $SO(1,1)$, we have again raised indices using $\eta$ and $\nabla_\mu$ is the Levi-Civita connection. To ensure closure of the gauge algebra the embedding tensors must satisfy the following algebraic constraints
\begin{align}\label{quadconsembed}
3f_{R[MN}f_{PQ]}{}^R&=2f_{[MNP}\xi_{Q]}\,,\qquad \xi_M{}^Qf_{QNP}=\xi_M\xi_{NP}-\xi_{[N}\xi_{P]M}\,,\nn
\xi_M\xi^M&=0\,,\qquad\xi_{MN}\xi^N=0\,,\qquad f_{MNP}\xi^P=0\,.
\end{align}

Associated with the vector fields $\mathcal{A}^0_{(1)}$ and $\mathcal{A}^M_{(1)}$, we also need to introduce
two-form gauge fields $\mathcal{B}_{(2)0}$ and $\mathcal{B}_{(2)M}$. In the ungauged theory these appear
on-shell as the Hodge duals of the fields strengths of the vectors. In the gauged theory the two-forms are introduced
as off-shell degrees of freedom, but the equations of motion ensure that the suitably defined covariant field strengths are still Hodge dual. In particular, the two-forms appear in the covariant field strengths for the vector fields, $\mathcal{H}^0_{(2)}$ and
$\mathcal{H}^M_{(2)}$, via
\begin{align}\label{tfmdefbee}
\mathcal{H}^M_{(2)}=&\,d\mathcal{A}_{(1)}^M-\tfrac{1}{2}g f_{NP}{}^M\mathcal{A}_{(1)}^N\wedge \mathcal{A}_{(1)}^P
-\tfrac{1}{2}g \xi_P{}^M\mathcal{A}_{(1)}^0\wedge \mathcal{A}_{(1)}^P
+\tfrac{1}{2}g \xi_P\mathcal{A}_{(1)}^M\wedge \mathcal{A}_{(1)}^P\nn
&
+\tfrac{1}{2}g \xi^{MN}\mathcal{B}_{(2)N}
-\tfrac{1}{2}g \xi^{M}\mathcal{B}_{(2)0}\,,\nn
\mathcal{H}^0_{(2)}=&\,
d\mathcal{A}_{(1)}^0+\tfrac{1}{2}g \xi_M\mathcal{A}_{(1)}^M\wedge \mathcal{A}_{(1)}^0
+\tfrac{1}{2}g \xi^{M}\mathcal{B}_{(2)M}\,.
\end{align}
The equations of motion are invariant under gauge transformations, with spacetime dependent parameters $(\Lambda^0,\Lambda^M)$. 
In addition there are gauge transformations parametrised by the 
spacetime dependent one-forms $(\Xi_{(1)0},\Xi_{(1)M})$ that just act on the one-forms and two-forms.
In particular, acting on these fields we have
\begin{align}\label{gt2fms}
\delta \mathcal{A}_{(1)}^M&=D\Lambda^M-\tfrac{1}{2}g\xi^{MN}\Xi_{(1)N}+\tfrac{1}{2}g\xi^{M}\Xi_{(1)0}\,,\nn
\delta \mathcal{A}_{(1)}^0&=D\Lambda^0-\tfrac{1}{2}g\xi^{M}\Xi_{(1)M}\,,\nn
\delta \mathcal{B}_{(2)M}&=D\Xi_{(1)M}-{2}\mathcal{H}^0_{(2)}\Lambda_M-{2}\mathcal{H}_{(2)M}\Lambda^0\,,\nn
\delta \mathcal{B}_{(2)0}&=D\Xi_{(1)0}-{2}\mathcal{H}_{(2)M}\Lambda^M\,.
\end{align}

With these ingredients in hand, the $N=4$ gauged supergravity
Lagrangian can be written as\footnote{Note that we have multiplied the
Lagrangian in \cite{Schon:2006kz} by a factor of two.} the five-form
\begin{align}\label{n4lag1}
\mathcal{L}_{N=4}=&\,{R}\mathrm{vol}_{5}+\mathcal{L}^S_{N=4}+\mathcal{L}^{pot}_{N=4}+\mathcal{L}^V_{N=4}+\mathcal{L}^T_{N=4}\,.
\end{align}
Here the scalar kinetic energy terms are given by
\begin{align}
\mathcal{L}^S_{N=4}=-3\Sigma^{-2}{\ast d\Sigma}\wedge d\Sigma
+\frac{1}{8}{\ast D \mathcal{M}_{MN}}\wedge D  \mathcal{M}^{MN}\,,
\end{align}
and the scalar potential is given by
\begin{align}\label{n4scalpot}
\mathcal{L}^{pot}_{N=4}=&-\tfrac{1}{2}{g}^2\Big\{f_{MNP}f_{QRS}\Sigma^{-2}\left(\tfrac{1}{12}\mathcal{M}^{MQ}\mathcal{M}^{NR}\mathcal{M}^{PS}-\tfrac{1}{4}\mathcal{M}^{MQ}\eta^{NR}\eta^{PS}+\tfrac{1}{6}\eta^{MQ}\eta^{NR}\eta^{PS}\right)\nn
&\phantom{-\tfrac{1}{2}{g}^2\Big\{}+\tfrac{1}{4}\xi_{MN}\xi_{PQ}\Sigma^4\Big(\mathcal{M}^{MP}\mathcal{M^{NQ}}-\eta^{MP}\eta^{NQ}\Big)+\xi_{M}\xi_{N}\Sigma^{-2}\mathcal{M}^{MN}\nn
&\phantom{-\tfrac{1}{2}{g}^2\Big\{}+\tfrac{1}{3}\sqrt{2}f_{MNP}\xi_{QR}\Sigma\mathcal{M}^{MNPQR}\Big\}\text{vol}_{5}\,.
\end{align}
The kinetic terms for the vectors,
which also involve two-form contributions via \eqref{tfmdefbee}, are given by
\begin{equation}\label{n4lag4}
\mathcal{L}^{V}_{N=4}=-\Sigma^{-4}{\ast \mathcal{H}^0_{(2)}}\wedge \mathcal{H}^0_{(2)}
-\Sigma^{2}\mathcal{M}_{MN} {\ast \mathcal{H}^M_{(2)}}\wedge \mathcal{H}^N_{(2)}\,.
\end{equation}
In order to succinctly present the topological part of the Lagrangian in \eqref{n4lag1}, we 
temporarily introduce the calligraphic index $\mathcal{M}=(0,M)$
which allows us to package the 9 vector fields and 9 two-forms into the quantities
$\mathcal{A}_{(1)}^\mathcal{M}$ and $\mathcal{B}_{(2)\mathcal{M}}$, each transforming in the fundamental representation of $SO(1,1)\times SO(5,3)$. In the conventions of this
paper\footnote{Throughout this paper
we take, in an orthonormal frame, $\epsilon_{01234}=+1$ so that $\epsilon=\mathrm{vol}_{5}$.
We have assumed that \cite{Schon:2006kz} have taken $\epsilon_{01234}=-1$ and then the expression for the topological term given here agrees with that in \cite{Schon:2006kz} up to an overall factor of 2.}, we then have
\begin{align}\label{n4lag5}
\mathcal{L}^{T}_{N=4}=&
-\frac{1}{\sqrt{2}}gZ^{\mathcal{M}\mathcal{N}}\mathcal{B}_{\mathcal{M}}\wedge D\mathcal{B}_{\mathcal{N}}-\sqrt{2}gZ^{\mathcal{M}\mathcal{N}}\mathcal{B}_{\mathcal{M}}\wedge d_{\mathcal{N}\mathcal{P}\mathcal{Q}}\mathcal{A}^{\mathcal{P}}\wedge d\mathcal{A}^{\mathcal{Q}}\nn
&-\frac{\sqrt{2}}{3}{g}^2 Z^{\mathcal{M}\mathcal{N}}\mathcal{B}_{\mathcal{M}}\wedge d_{\mathcal{N}\mathcal{P}\mathcal{Q}}\mathcal{A}^{\mathcal{P}}\wedge X_{\mathcal{R}\mathcal{S}}^{\phantom{\mathcal{R}\mathcal{S}}\mathcal{Q}}\mathcal{A}^{\mathcal{R}}\wedge \mathcal{A}^{\mathcal{S}}+\frac{\sqrt{2}}{3}d_{\mathcal{M}\mathcal{N}\mathcal{P}}\mathcal{A}^{\mathcal{M}}\wedge d\mathcal{A}^{\mathcal{N}}\wedge d\mathcal{A}^{\mathcal{P}}\nn
&+\frac{1}{2\sqrt{2}}{g}d_{\mathcal{M}\mathcal{N}\mathcal{P}}X_{\mathcal{Q}\mathcal{R}}^{\phantom{\mathcal{Q}\mathcal{R}}\mathcal{M}}\mathcal{A}^{\mathcal{N}}\wedge \mathcal{A}^{\mathcal{Q}}\wedge \mathcal{A}^{\mathcal{R}}\wedge d\mathcal{A}^{\mathcal{P}}\nn
&+\frac{1}{10\sqrt{2}}{g}^2d_{\mathcal{M}\mathcal{N}\mathcal{P}}X_{\mathcal{Q}\mathcal{R}}^{\phantom{\mathcal{Q}\mathcal{R}}\mathcal{M}}X_{\mathcal{S}\mathcal{T}}^{\phantom{\mathcal{Q}\mathcal{R}}\mathcal{P}}\mathcal{A}^{\mathcal{N}}\wedge \mathcal{A}^{\mathcal{Q}}\wedge \mathcal{A}^{\mathcal{R}}\wedge \mathcal{A}^{\mathcal{S}}\wedge \mathcal{A}^{\mathcal{T}}\,.
\end{align}
Here the symmetric tensor $d_{\mathcal{M}\mathcal{N}\mathcal{P}}=d_{(\mathcal{M}\mathcal{N}\mathcal{P})}$ has non-zero components
\begin{align}
d_{0MN}=d_{M0N}=d_{MN0}=\eta_{MN}\,,
\end{align}
the antisymmetric tensor 
$Z^{\mathcal{M}\mathcal{N}}=Z^{[\mathcal{M}\mathcal{N}]}$ has components
\begin{align}
Z^{MN}=\tfrac{1}{2}\xi^{MN}\,,\qquad
Z^{0M}=-Z^{M0}=\tfrac{1}{2}\xi^{M}\,,\qquad
\end{align}
and the only non-zero components of $X_{\mathcal{M}\mathcal{N}}{}^{\mathcal{P}}$ are given by
\begin{align}
X_{MN}{}^P=-f_{MN}{}^P-\tfrac{1}{2}\eta_{MN}\xi^P+\delta^P_{[M}\xi_{N]}\,,\quad
X_{M0}{}^0=\xi_M\,,\quad
X_{0M}{}^N=-\xi_M{}^N\,.
\end{align}
It is worth noting that after defining the matrices $(X_{\mathcal{M}})_\mathcal{N}{}^{\mathcal{P}}\equiv X_{\mathcal{M}\mathcal{N}}{}^{\mathcal{P}}$ we have $[X_\mathcal{M},X_\mathcal{N}]=-
X_{\mathcal{M}\mathcal{N}}{}^{\mathcal{P}}X_\mathcal{P}$, by virtue of the quadratic constraints satisfied by the embedding tensor given in \eqref{quadconsembed}.

Shortly it will be useful to note that 
the two forms only appear in the Lagrangian in one of the following two combinations
\begin{align}\label{tfcombs}
\xi^{MN}\mathcal{B}_{(2)N}-\xi^M \mathcal{B}_{(2)0}\,,\qquad \xi^N\mathcal{B}_{(2)N}\,.
\end{align}

\subsection{Matching}
We now match the $D=5$ theory of section \ref{sec3} with the $N=4$ gauged theory presented in the previous subsection. We first discuss the scalar field sector
and then subsequently discuss the gauging and the embedding tensor.

\subsubsection{Identifying the scalar fields}
We take
the generators of $SO(5,3)$ to be given by the 
$8\times 8$ matrices\footnote{Note that this differs by a factor of two compared with \cite{Schon:2006kz} as mentioned in footnote \ref{foot2}.}
\begin{equation}\label{explicgen}
(t_{MN})^A\,_B = \delta^A_M\eta_{BN}-\delta^A_N\eta_{MB}\,,
\end{equation}
with $\eta$, non-diagonal, as in \eqref{etadef}. 
In order to parametrise the coset $SO(5,3)/(SO(5)\times SO(3))$ we exponentiate a suitable solvable subalgebra of the
Lie algebra. Following, for example \cite{Lu:1998xt}, the three non-compact Cartan generators $H^i$ and the twelve positive root
generators, with positive weights under $H^i$,  
are given by\footnote{To compare with (3.31) of \cite{Lu:1998xt} we should make the identifications
$(T^1,T^2,T^3)=(E_1{}^2,E_1{}^3,E_2{}^3)$,
$(T^4,T^5,T^6)=(V^{12},V^{13},V^{23})$,
$(T^7,T^8,T^9)=(U_1^1,U_1^2,U_1^3)$ and
$(T^{10},T^{11},T^{12})=(U_2^1,U_2^2,U_2^3)$.
}
\begin{align}\label{eq:generators}
&H^{1}=\sqrt{2}t_{16}\,, \quad H^{2}=\sqrt{2}t_{27}\,,\quad H^{3}=\sqrt{2}t_{38}\,,\nn
&T^{1}=-t_{26}\,,\quad T^{2}=-t_{36}\,,\quad T^{3}=-t_{37}\,, \quad T^{4}=t_{12}\,,\quad T^{5}=t_{13}\,,\quad T^{6}=t_{23}\,,\nn
&T^{7}=-t_{14}\,,\,\,T^{8}=-t_{24}\,, \,\,T^{9\phantom{0}}=-t_{34}\,,\,\,
T^{10}=-t_{15}\,,\,\,T^{11}=-t_{25}\,,\, \,T^{12}=-t_{35}\,.
\end{align}
We note that $\text{Tr}(T^i(T^j)^T)=2\delta^{ij}$ and $\text{Tr}(H^mH^n)=4\delta^{mn}$ with $H^m=(H^m)^T$.

To make contact with the scalar fields in the reduced equations of motion of section \ref{sec3}, 
we first need an explicit embedding of the coset 
$SL(3)/SO(3)$ inside $SO(5,3)/(SO(5)\times SO(3))$. This is conveniently achieved 
by first defining 
\begin{align}\label{slthem}
\mathscr{H}^1&=H^2-H^1\,,\quad \mathscr{H}^2=H^3-H^2\,,\quad \mathcal{E}^{1}=T^1\,, \quad  \mathcal{E}^{2}=T^3\,,\quad  \mathcal{E}^{3}=T^2\,,
\end{align}
as well as $\mathscr{H}^3=-(H^1+H^2+H^3)$ which commutes with all five of the generators in \eqref{slthem}. 
By introducing six scalar fields $\varphi_i$ and $a_i$ we can consider the coset element
\begin{align}
\mathcal{V}_{(S)}&=e^{\tfrac{1}{\sqrt{2}}\vec{\varphi}\cdot \vec{\mathscr{H}}}e^{a_1\mathcal{E}^1}e^{a_2\mathcal{E}^2}e^{a_3\mathcal{E}^3}\,,\nn
&=\left(
\begin{array}{ccc}
 e^{-\varphi _3}V^{-T}&0 &0 \\
 0 & \mathds{1}_{2\times2} & 0 \\
 0 & 0 & e^{\varphi _3}V \\
\end{array}
\right)\,,
\end{align}
where the $3\times 3$ matrix $V$ parametrises the coset $SL(3)/SO(3)$ in a standard upper triangular gauge (see appendix \ref{appc}):
\begin{align}\label{sl3co3coset}
V=\left(
\begin{array}{ccc}
 e^{\varphi _1} & e^{\varphi _1} a_1 & e^{\varphi _1} \left(a_1 a_2+a_3\right) \\
 0 & e^{\varphi _2-\varphi _1} & e^{\varphi _2-\varphi _1} a_2 \\
 0 & 0 & e^{-\varphi _2} \\
\end{array}
\right)\,.
\end{align}
Moreover, we can identify the scalar fields in the $3\times 3$ matrix 
$\mathcal{T}^{\alpha\beta}$ in the reduced theory of section \ref{sec3}
via
\begin{align}
\mathcal{T}^{\alpha\beta}=(V^TV)^{\alpha\beta}\,.
\end{align}
As already anticipated in \eqref{scalidents}, we next note that the scalar field $\Sigma$, that parametrises $SO(1,1)$ in the $N=4$ theory and the 
scalar field $\varphi_3$ can be identified with the scalar fields $\phi,\lambda$ in the reduced theory
of section \ref{sec3} via
\begin{equation}
\varphi_3 = 3\phi-\lambda\,,\qquad \Sigma = e^{-(\phi+3\lambda)}\,.
\end{equation}

Having clarified this embedding we next define the coset element, $\mathcal{V}$, which parametrises $SO(5,3)/(SO(5)\times SO(3))$ and incorporates the remaining scalars
$\xi^\alpha$ and $\psi^{a\alpha}$ of section \ref{sec3}, via
\begin{align}\label{cosettext}
\mathcal{V}=&\,\mathcal{V}_{(S)} e^{(\xi^3-\psi^{a1}\psi^{a2})T^4}
e^{-(\xi^2+\psi^{a3}\psi^{a1})T^5}
e^{(\xi^1-\psi^{a2}\psi^{a3})T^6}\nn
&\qquad\qquad\cdot e^{\sqrt{2}\psi^{11}T^7}
e^{\sqrt{2}\psi^{12}T^8}e^{\sqrt{2}\psi^{13}T^9}e^{\sqrt{2}\psi^{21}T^{10}}e^{\sqrt{2}\psi^{22}T^{11}}e^{\sqrt{2}\psi^{23}T^{12}}\,.
\end{align}

\subsubsection{The Embedding tensor}
We claim that the reduced $D=5$ theory of section \ref{sec3} is an $N=4$ gauged supergravity
with gauge group $SO(2)\times SE(3) \subset SO(5,3)$, where 
$SE(3)$ is the three-dimensional special Euclidean group. The compact $SO(2)\times SO(3)$ subgroup is generated by
\begin{equation}
\mathfrak{g}_0=t_{45}\,, \quad \mathrm{and} \quad 
\mathfrak{g}_1=t_{37}-t_{28}\,, \quad \mathfrak{g}_2=-(t_{36}-t_{18})\,, \quad \mathfrak{g}_3=t_{26}-t_{17}\,,
\end{equation}
with e.g. $[\mathfrak{g}_1,\mathfrak{g}_2]=\mathfrak{g}_3$ 
and the additional non-compact generators in $SE(3)$ are given by
\begin{equation}
\mathfrak{g}_4=t_{23}\,,\quad \mathfrak{g}_5=-t_{13}\,,\quad \mathfrak{g}_6=t_{12}\,.
\end{equation}
The components of the embedding tensor are specified by\footnote{If we use \eqref{usimmat}
to move to a basis in which $\eta_{MN}$ is diagonal, then the independent components
are given by $\bar{f}_{123} = -\frac{1}{2}(3+l)$, $\bar{f}_{678} = \frac{1}{2}(3-l)$, 
$\bar{f}_{128} =\bar{f}_{236}=-\bar{f}_{137}= -\frac{1}{2}(l+1)$ and
$\bar{f}_{178} = -\bar{f}_{268}=\bar{f}_{367}=\frac{1}{2}(1-l)$. 
We also note that since $\xi^M=0$, the gauged supergravity lies within the class constructed in \cite{DallAgata:2001wgl}.}
\begin{align}\label{embeddingoef}
\xi^M=0\,,\qquad\qquad\quad
\xi^{45}=-\sqrt{2}\,,\nn
f_{187} = f_{268} = f_{376} = {\sqrt{2}}\,, \qquad\qquad f_{678} = {l}{\sqrt{2}}\,,
\end{align}
along with the fact that $f_{MNP}=f_{[MNP]}$, $\xi^{NP}=\xi^{[NP]}$ and the remaining components are all zero.

With this specific embedding tensor, we can make two important simplifications to
the $N=4$ theory. First, since the two-forms only appear in the combinations given by
\eqref{tfcombs}, we can set the following components to zero
\begin{align}
\mathcal{B}_{(2)0}=0\,,\qquad\mathcal{B}_{(2)M=\alpha}=0\,,\qquad\mathcal{B}_{(2)M=5+\alpha}=0\,,
\end{align}
for $\alpha=1,2,3$. Second, we can use the gauge transformations given
in \eqref{gt2fms}, with parameters $\Xi_{(1)M=4},\Xi_{(1)M=5}$ to set the following
components of the gauge fields to zero
\begin{align}
\mathcal{A}_{(1)M=4}=0\,,\quad\mathcal{A}_{(1)M=5}=0\,.
\end{align}
Having done this we can identify the remaining gauge fields and two-forms of the $N=4$ theory with those
of the reduced theory given in section \ref{sec3} via
\begin{align}
\mathcal{A}^{0}_{(1)}=\tfrac{1}{\sqrt{2}}A_{(1)}\,,\qquad
\mathcal{A}^{M=\alpha}_{(1)}=\tfrac{1}{\sqrt{2}}(\mathscr{A}_{(1)}^\alpha-lA^\alpha_{(1)})\,,
\qquad\mathcal{A}^{M=5+\alpha}_{(1)}=-\tfrac{1}{\sqrt{2}}A^\alpha_{(1)}\,,
\end{align}
with $\alpha=1,2,3$ (and recalling \eqref{relabalphind}) as well as 
\begin{align}
\mathcal{B}^{4}_{(2)}=\frac{1}{g}L^2_{(2)}\,,\qquad
\mathcal{B}^{5}_{(2)}=-\frac{1}{g}L^1_{(2)}\,.
\end{align}
In particular, the covariant two-form field strengths of the $N=4$ theory given
in \eqref{tfmdefbee} are related to those of the reduced theory in section \ref{sec3} via
\begin{align}\label{eq.vector_identification}
\mathcal{H}^0_{(2)}=\tfrac{1}{\sqrt{2}}F_{(2)}\,,\qquad\mathcal{H}^M_{(2)}=\tfrac{1}{\sqrt{2}}(G^\alpha_{(2)},{L}^a_{(2)},-{F}^{\alpha}_{(2)})\,.
\end{align}
Furthermore, the covariant derivative in \eqref{eq_cov_derivative} is given by
\begin{equation}\label{eq_cov_derivative2}
D_\mu=\nabla_\mu+g\left(A_{\mu}\mathfrak{g}_0+A^1_{\mu}\mathfrak{g}_1+A^2_{\mu}\mathfrak{g}_2+A^3_{\mu}\mathfrak{g}_3+\mathscr{A}^1_{\mu}\mathfrak{g}_4+\mathscr{A}^2_{\mu}\mathfrak{g}_5+\mathscr{A}^3_{\mu}\mathfrak{g}_6\right)\,.
\end{equation}

With the above identifications of the fields and the given embedding tensor, one can show 
that the Lagrangian of the $D=5$ theory given in \eqref{lagfred1}-\eqref{eq:new_top_after_field_redef} is precisely equivalent to the $N=4$ Lagrangian
given in \eqref{n4lag1}-\eqref{n4lag5}. We have presented a few details of this calculation in appendix \ref{appc}.

\section{Consistent subtruncations}\label{subtrunc}

In this section we explore various consistent subtruncations of the reduced equations of motion given 
in \eqref{redeq1}-\eqref{redeq2} and \eqref{redeq3}-\eqref{redeq10}.

\subsection{Romans' $D=5$ $SU(2)\times U(1)$ supergravity theory}\label{romans}
When $l=-1$ ({\it i.e.} $\Sigma_2=H^2$), we can recover the Romans' $D=5$ $SU(2)\times U(1)$ gauged supergravity theory, maintaining half
maximal supersymmetry.
The fact that this must be possible immediately follows from the results of \cite{Gauntlett:2007sm}.

Specifically, we take
\begin{align}
l=-1\,,\qquad
\lambda =3\phi\,,
\end{align}
and set all of the remaining scalar fields to their trivial values
$\mathcal{T}_{\alpha\beta}=\delta_{\alpha\beta}$, $\db^{a\alpha}=0$.
We keep the two-forms and package them into a complex two-form via
\begin{equation}
\mathcal{C}_{(2)}=\K^1_{(2)}+i\K^2_{(2)}\,.
\end{equation}
Finally, we set $\chi^\alpha_{(1)}=0$ and impose 
\begin{align}
{\ast h^\alpha_{(3)}}=\tfrac{1}{2}e^{-20\phi}\epsilon_{\alpha\beta\gamma}{F}^{\beta\gamma}_{(2)}\,.
\end{align}
The field content now consists of a metric, a scalar field $\phi$, $SO(2)\times SO(3)\simeq U(1) \times SU(2)$ 
gauge fields $A_{(1)}$, $A^{\alpha\beta}_{(1)}$ and a complex two-form $\mathcal{C}_{(2)}$ which is charged 
under the $U(1)$ gauge field.
The truncated equations of motion are given in \eqref{romanseq1},\eqref{romanseq2} 
and are precisely\footnote{For example, we can compare with section 2.2. of \cite{Gauntlett:2007sm} by making the identifications
$\tfrac{1}{2}\epsilon_{\alpha\beta\gamma}A_{(1)}^{\beta\gamma}\to -2^{-1/6} A^\alpha$, $A_{(1)}\to 2^{-2/3} B$, $\mathcal{C}_{(2)}\to 2^{-1/6}C$, $e^{10\phi}\to 2^{1/3} X$ and $g\to -2^{2/3}m$.} 
 that of Romans' theory \cite{Romans:1985ps}
coming from the Lagrangian
\begin{align}\label{roamnstextlag}
\mathcal{L}^{Romans}=&\,{R}\mathrm{vol}_{5}-300{\ast d\phi}\wedge{d\phi}-\tfrac{1}{2}e^{40\phi}{\ast F_{(2)}}\wedge F_{(2)}
-\tfrac{1}{2}e^{-20\phi}{\ast {F}^{\alpha\beta}_{(2)}}\wedge {F}^{\alpha\beta}_{(2)}\nn
&-e^{-20\phi}{\ast \overline{\mathcal{C}}_{(2)}}\wedge \mathcal{C}_{(2)}
+\tfrac{1}{2ig}\left(\overline{\mathcal{C}}_{(2)}\wedge D\mathcal{C}_{(2)}-\mathcal{C}_{(2)}\wedge D\overline{\mathcal{C}}_{(2)}\right)
\nn
&+{g}^2(4e^{-10\phi}+e^{20\phi})\text{vol}_5-\tfrac{1}{2}{F}^{\alpha\beta}_{(2)}\wedge {F}^{\alpha\beta}_{(2)}\wedge A_{(1)}\,,
\end{align}
and $D\mathcal{C}_{(2)}=d\mathcal{C}_{(2)}-igA_{(1)}\wedge \mathcal{C}_{(2)}$.
We note that this Lagrangian can also be obtained by directly substituting the ansatz into the $D=5$ Lagrangian.

As is well known we can then further truncate Romans' theory to minimal $N=2$, $D=5$ gauged supergravity. In the notation here,
this can be achieved by imposing
$e^{10 \phi}=2^{1/3}$, setting the two-forms to zero, $\mathcal{C}_{(2)}=0$, and keeping a single $U(1)$ gauge field in the diagonal of
$U(1)\times SU(2)$ via ${F}^{12}_{(2)}=2F_{(2)}$ and  ${F}^{23}_{(2)}={F}^{31}_{(2)}=0$. The
resulting equations of motion arise from the Lagrangian for minimal gauged supergravity given by
\begin{align}
\mathcal{L}^{Min}=&\,{R}\mathrm{vol}_{5}-3\cdot2^{1/3}\, {\ast F_{(2)}}\wedge F_{(2)}
+3\cdot2^{2/3}\, {g}^2\text{vol}_5
-4F_{(2)}\wedge F_{(2)}\wedge A_{(1)}\,.
\end{align}

It is worth emphasising that these two subtruncations cannot exist when $l=1,0$,
({\it i.e.} $\Sigma_2=S^2,\mathbb{R}^2$). Indeed, if they did exist, then the maximally supersymmetric solution
of these theories would necessarily be associated with a maximally supersymmetric $AdS_5$ solution of the $N=4$, $D=5$ gauged supergravity theory, which do not exist, as we show in section \ref{susvacsec}.

\subsection{Various invariant sectors}
There are various additional truncations, for all cases $l=0,\pm1$, that arise from keeping sectors invariant under various subgroups of 
$SO(2)\times SO(3)$.

\subsubsection{$SO(3)$ invariant sector}
A simple truncation is to keep only the fields that transform as singlets under $SO(3)$. Setting
$h_{(3)}^\alpha=\chi_{(1)}^\alpha=\psi^{a\alpha}=A^{\alpha\beta}=0$ and 
$\mathcal{T}^{\alpha\beta}=\delta^{\alpha\beta}$ in
the $D=5$ equations of motion \eqref{redeq1}-\eqref{redeq2} and \eqref{redeq3}-\eqref{redeq10}
leads to a consistent set of equation of motion. The fields kept in this truncation
consist of the metric as well as
\begin{align}
\phi\,,\lambda \,,A_{(1)}\,,K^a_{(2)}\,.
\end{align}
It is consistent with the equations of motion to further set the two-forms to zero $K^a_{(2)}=0$.
We note that this truncation cannot be further truncated to minimal gauged supergravity.

\subsubsection{$SO(2)_R\subset SO(3)$ invariant sector}
We can slightly extend the truncation just considered, by keeping fields that are invariant
under a subgroup $SO(2)_R\subset SO(3)$. More specifically, we consider an $SO(3)$ triplet, 
with index $\alpha=1,2,3$  to decompose into a doublet and a  
singlet of $SO(2)_R$, with indices
$\alpha=1,2$ and $\alpha=3$, respectively. The fields that are kept in this truncation are the metric and
\begin{align}\label{so2ltrunc}
\phi\,,\lambda\,, A_{(1)}\,, K^a_{(2)}\,,
\mathcal{T}_{\alpha\beta}=\text{diag}(e^{w},e^{w},e^{-2w})\,,\psi^{a3}\,,A^{12}_{(1)}\,,\chi_{(1)}^3\,,h^3_{(3)}\,.
\end{align}

\subsubsection{$SO(2)$ invariant sector}
We can also consider the truncation that keeps the fields that are invariant under the explicit $SO(2)$ factor in
$SO(2)\times SO(3)$. The fields that 
are kept in this truncation are the metric and
\begin{align}
\phi\,,\lambda\,, \mathcal{T}_{\alpha\beta}\,,
A_{(1)}\,,A^{\alpha\beta}_{(1)}\,,
\chi_{(1)}^\alpha\,, h^\alpha_{(3)}\,.
\end{align}

\subsection{Diagonal $SO(2)_D$ invariant sector}\label{so2dsec}
The final subtruncation we consider, again for all cases $l=0,\pm1$, keeps the sector that is invariant under 
an $SO(2)_D$ diagonal subgroup of $SO(2)\times SO(2)_R\subset SO(2)\times SO(3)$
where $SO(2)_R\subset SO(3)$ was defined
in the previous subsection. This is a particularly interesting truncation since we show that
it is consistent
with $N=2$ supersymmetry. Specifically we show that we obtain the bosonic sector of an $N=2$,
$D=5$  gauged
supergravity coupled to two vector multiplets, with the two scalars parametrising the very special real manifold 
$SO(1,1)\times SO(1,1)$, and a single hypermultiplet, with the four scalars parametrising the quaternionic 
manifold $SU(2,1)/S[U(2)\times U(1)]$. Furthermore, the gauging is just in the hypermultiplet sector.

In restricting to the $SO(2)_D$ invariant fields we should set $\psi^{a3}=\K^a_{(2)}=0$
in \eqref{so2ltrunc} but we can 
now keep an additional two scalar modes in the $\psi^{a\alpha}$ sector with $\alpha=1,2$, specifically, 
\begin{align}
z^1\equiv \tfrac{1}{2}(\psi^{11}+\psi^{22})\,,\qquad
z^2\equiv \tfrac{1}{2}(\psi^{21}-\psi^{12})\,.
\end{align}
This can be achieved by imposing
\begin{align}\label{psitrcgh}
\psi^{a2}=-\epsilon_{ab}\psi^{b1}\,,
\end{align}
and keeping the fields 
\begin{equation}
\phi\,,\lambda\,, \mathcal{T}_{\alpha\beta}=\text{diag}(e^{w},e^{w},e^{-2w})\,,z^a\,,
A_{(1)}\,,A^{12}_{(1)}\,,\chi_{(1)}^3\,, h^3_{(3)}\,,
\end{equation}
as well as the metric.
Note that using \eqref{psitrcgh} we have $z^1=\psi^{11}$, $z^2=\psi^{21}$. Furthermore,
the covariant derivative acting on $z^a$ and the field strengths are now given by
\begin{equation}\label{newcdv}
F_{(2)}=dA_{(1)}\,,\quad {F}^{12}_{(2)}=dA^{12}_{(1)}\,,\quad Dz^a=dz^a+
g\epsilon_{ab}(-A^{12}_{(1)}+A_{(1)})z^b\,,
\end{equation}
and we notice that $z^a$, which is a singlet with respect to the diagonal $SO(2)$, 
is a doublet of the anti-diagonal $SO(2)$.
It is straightforward to show that this is a consistent truncation of
the $D=5$ equations of motion \eqref{redeq1}-\eqref{redeq2} and \eqref{redeq3}-\eqref{redeq10}. 

To display the $N=2$ structure of the truncated theory, it is convenient, as in section \ref{fieldredef}, to carry out some field redefinitions. 
We re-define $\chi^3_{(1)}$ and  $h^3_{(3)}$ into $\xi$ and $\mathscr{A}_{(1)}$ in the following way,
\begin{align}
\chi^3_{(1)}&\equiv d\xi+g\mathscr{A}_{(1)}-2\epsilon_{ab}z^aDz^b\,,\nn
{* h^3_{(3)}}&\equiv e^{-4\lambda-8\phi+2w}G_{(2)}\,,
\end{align}
where
\begin{equation}
G_{(2)}\equiv d(\mathscr{A}_{(1)}-lA^{12}_{(1)})\,,
\end{equation}
and one can check that these redefinitions are consistent with the equations of motion.
We also replace the three scalar fields $\{\phi,\lambda,w\}$ with $\{\Sigma,\Om,\varphi\}$ defined as
\begin{equation}
\Sigma = e^{-(\phi+3\lambda)}\,,\qquad \Om=e^{3\phi-\lambda-w}\,,\qquad
\varphi=\lambda-3\phi-\tfrac{1}{2}w\,.
\end{equation}
After substituting these redefinitions into the equations of motion, we find equations of motion that can be
derived from the action with Lagrangian
\begin{align}\label{afredfnrwo}
\mathcal{L}=&\,{R}\mathrm{vol}_{5}
-\tfrac{1}{2}\Sigma^{-4}{* F_{(2)}}\wedge F_{(2)}
-\tfrac{1}{2}\Sigma^2 \Om^2{* {F}^{12}_{(2)}}\wedge {F}^{12}_{(2)}
-\tfrac{1}{2}\Sigma^2 \Om^{-2}{* G_{(2)}}\wedge G_{(2)}\nn
&
-3\Sigma^{-2}\ast d\Sigma\wedge{d\Sigma}
-\Om^{-2}{\ast d\Om}\wedge{d \Om}
- A_{(1)} \wedge{F}^{12}_{(2)}\wedge G_{(2)}\nn
&-{2}{* d\varphi}\wedge d\varphi
-\tfrac{1}{2}e^{4\varphi}{* (d\xi+g\mathscr{A}_{(1)}-2\epsilon_{ab}z^{a}Dz^{b}})\wedge (d\xi+g\mathscr{A}_{(1)}-2\epsilon_{cd}z^{c}Dz^{d})\nn
&-2e^{2\varphi}{* Dz^{a}}\wedge Dz^{a}\nn
&+{g}^2\Om^{-2}\Sigma^{-2}\Big\{ 2le^{2\varphi}\Om\Sigma^3-\tfrac{1}{2}e^{4\varphi}(l-2 z^az^a)^2     
-{2}e^{4\varphi}\Om^2\Sigma^6 (z^az^a)^2          \nn
&-\tfrac{1}{2}e^{4\varphi}\Om^4+4\Om\Sigma^3+2e^{2\varphi}\Om^2+2e^{2\varphi}\Om^3\Sigma^3
-{2}e^{2\varphi}(1-\Om\Sigma^3)^2z^az^a\Big\}\mathrm{vol}_{5}\,.
\end{align}

We now recall a general class of $N=2$, $D=5$ gauged supergravity theories that are
coupled to two vector multiplets and 
a single hypermultiplet, following \cite{Bergshoeff:2004kh} (which generalised \cite{Gunaydin:1984ak,Gunaydin:1999zx,Ceresole:2000jd,Gunaydin:2000ph}). The Lagrangian for the bosonic fields
can be written
\begin{align}\label{bosntwgen}
\mathcal{L}_{N=2}={R}\mathrm{vol}_{5}&-\tfrac{1}{2}a_{IJ}{\ast H^{I}}\wedge H^{J}-\tfrac{1}{2}g_{xy}{\ast D\phi^{x}}\wedge D\phi^y-\tfrac{1}{3\sqrt{3}}\mathcal{C}_{IJK}A^I\wedge F^J\wedge F^K\nn
&
-\tfrac{1}{2}g_{XY}{\ast Dq^X}\wedge Dq^Y
+\mathcal{L}^{pot}_{N=2}\,,
\end{align}
where the scalar potential $\mathcal{L}^{pot}_{N=2}$ is written in appendix \ref{appd} and 
\begin{align}\label{ntwocder}
D\phi^x\equiv d\phi^x+gA^{I}K^x_{I}\,,\, Dq^X\equiv d q^X+gA^{I}k^X_{I}\,,\, H^I\equiv dA^I+\tfrac{1}{2}g\bar f_{JK}^{\phantom{JK}I}A^J\wedge A^K\,.
\end{align}
Here $A^I$, with $I=0,1,2$, label the graviphoton as well as the two vector fields in the two vector multiplets and 
$\phi^x$, with $x,y=1,2$, are the associated two real scalar fields that parametrise a two dimensional very special real manifold which we take to be $SO(1,1)\times SO(1,1)$.
The $q^X$, with $X=1,\dots,4$, are the four real scalar fields in the hypermultiplet that parametrise a quaternionic K\"ahler space, which must be $SU(2,1)/S[U(2)\times U(1)]$. 
In the covariant derivatives $K^x_{I}$ and $k^X_{I}$ are each a set of three Killing vectors on the 
very special real manifold and on the quaternionic K\"ahler manifold, respectively. 
The structure constants of the gauge group are
given by $\bar f_{JK}^{\phantom{JK}I}$.
We now explain how our truncated Lagrangian \eqref{afredfnrwo} can be cast in this form with gauging only
in the hypermultiplet sector, which moreover is abelian with $\bar f_{JK}^{\phantom{JK}I}=0$.

We start with the vector multiplets. The very special real geometry is determined by a real, symmetric, constant 
tensor $\mathcal{C}_{IJK}$ which specifies
the embedding of $SO(1,1)\times SO(1,1)$ in a three-dimensional space with coordinates $h^I$ via
\begin{align}
\mathcal{C}_{IJK}h^I h^J h^K=1\,.
\end{align}
Defining $h_I=\mathcal{C}_{IJK}h^J h^K$ we can define $a_{IJ}$, which provides the kinetic terms for the vectors in 
\eqref{bosntwgen}, via
\begin{align}\label{aijmet}
a_{IJ}=-2\mathcal{C}_{IJK}h^K+3h_I h_J\,.
\end{align}
Indices can be lowered and raised using $a_{IJ}$ and its inverse $a^{IJ}$, and we note in particular that
$h_I=a_{IJ}h^J$. Moreover, the pull-back of $a_{IJ}$ gives the metric for the scalar fields $\phi^x$
via
\begin{align}\label{gxymet}
g_{xy}=3\partial_x h^I\partial_y h^Ja_{IJ}\,.
\end{align}

With these definitions in hand we return to the truncated Lagrangian \eqref{afredfnrwo}. 
We see that $\Sigma,\Om$ parametrise $SO(1,1)\times SO(1,1)$ with 
\begin{align}
\mathcal{C}_{012}=\frac{\sqrt{3}}{2}\,,\qquad
h^I=\frac{1}{\sqrt{3}}(\Sigma^{2},-\Sigma^{-1} \Om^{-1},-\Sigma^{-1} \Om)\,,
\end{align}
and we can identify the vector fields as follows:
\begin{equation}
A^{I}=(A_{(1)},A^{12}_{(1)},\mathscr{A}_{(1)}-lA^{12}_{(1)})\,.
\end{equation}
It is then straightforward to show that the first two lines in \eqref{afredfnrwo} are precisely the
same form as the first two lines of the $N=2$, $D=5$ Lagrangian in \eqref{bosntwgen}.

We next turn to the hypermultiplet. From the third and fourth lines of the truncated Lagrangian
\eqref{afredfnrwo}, we identify the 
coordinates on the quaternionic K\"ahler manifold to be
\begin{align}
q^X=(\varphi,\xi,z^1,z^2)\,,
\end{align}
with associated metric 
\begin{align}\label{quatmet}
g_{XY}dq^Xdq^Y=\,4d\varphi^2+4e^{2\varphi}{dz^a} dz^a+e^{4\varphi}(d\xi-2\epsilon_{ab} z^a dz^b)^2\,.
\end{align}
This is indeed the homogeneous metric on $SU(2,1)/S[U(2)\times U(1)]$ as we explain in appendix \ref{appd}.
This metric includes Killing vectors $\partial_\xi$ and $z^2\partial_1-z^1\partial_2$, which generate an $SO(2)\times\mathbb{R}$ subgroup of $SU(2,1)$. The Killing vectors, $k_I^X$, that determine the gauging
in \eqref{ntwocder} are given by the following linear combinations
\begin{equation}\label{kvecsfin}
k_0=z^2\partial_1-z^1\partial_2\,,\qquad
k_1=l\partial_\xi+z^1\partial_2-z^2\partial_1\,,
\qquad
k_2=\partial_\xi\,.
\end{equation}

To conclude the discussion on supersymmetry, it remains to check that the scalar potential terms given in the last two lines of the
truncated Lagrangian
\eqref{afredfnrwo} coincide with $\mathcal{L}^{pot}_{N=2}$ in \eqref{bosntwgen}. We successfully carry out 
this check in appendix \ref{appd}.

Finally, we note that if we further consistently truncate the theory in \eqref{afredfnrwo} by setting the scalars $z^a$ that are charged under the $SO(2)$
gauge group to zero, as well as use the non-compact $\mathbb{R}$ gauge transformations to set the Stueckelberg scalar $\xi$ to zero, then we obtain
\footnote{For example, we can identify the scalar fields here with those in \cite{Szepietowski:2012tb} via
$\Sigma=e^{-B/3+\lambda_1}$, $\Omega=e^{B-\lambda_1-2\lambda_2}$ and $\varphi=-(B+\lambda_1+\lambda_2)$. We should also set
$p_2=0$ in \cite{Szepietowski:2012tb}.} a $D=5$ theory which was first constructed in \cite{Szepietowski:2012tb}. Thus,
the Lagrangian \eqref{afredfnrwo} comprises the $N=2$ supersymmetric completion of the $D=5$ theory of \cite{Szepietowski:2012tb}, the existence of which was also conjectured in \cite{Szepietowski:2012tb}.

\section{Some solutions of the $D=5$ theory}\label{secsols}

\subsection{Maximally supersymmetric $AdS_5$ vacuum}\label{susvacsec}
The maximally supersymmetric $AdS_5$ vacuum solution is obtained by setting $l=-1$, taking
\begin{equation}\label{sussol1}
e^{30\phi}=2\,,\qquad
e^{10\lambda}=2\,,
\end{equation}
with all other fields trivial, and the $AdS_{5}$ radius squared $L^2$ is given by
\begin{equation}\label{sussol2}
{g}^2 L^2=2^{4/3}\,.
\end{equation}
By uplifting this solution to $D=7$ and then to $D=11$, it is straightforward to see that this is the same $AdS_5$ solution that was constructed in
\cite{Maldacena:2000mw} that is associated with M5-branes wrapping a Riemann surface in a Calabi-Yau two-fold. In particular, 
the presence of the spin connection $\bar\omega^{ab}$ of the Riemann surface in
\eqref{gfans} precisely corresponds to the topological twist associated with such wrapped M5-branes.

Within the $N=4$, $D=5$ gauged supergravity theory, 
it is interesting to analyse the mass spectrum of the linearised perturbations of the fields about this supersymmetric vacuum.
The $\phi$,$\lambda$ equations of motion are coupled and gives rise to two scalars with 
$m^2L^2 =-4,12$ and are holographically dual to scalar operators with $\Delta=2,6$. The linearised scalars
in $\mathcal{T}$ are massless and are dual to operators with $\Delta=4$.
The six scalars $\db^{a\alpha}$ each have
have $m^2L^2 =5$ and are associated with scalar operators with $\Delta=5$. The two two-forms $\K^a_{(2)}$ give rise to operators with
$\Delta=3$. The vector $A_{(1)}$ is dual to a conserved current with $\Delta=3$ and the metric is dual to the stress tensor with $\Delta=4$.
A little work is required to decouple the linearised $h_{(3)}^\alpha,\chi_{(1)}^\alpha,F^{\alpha\beta}_{(2)}$ sector. One can first solve the linearised equation
\eqref{redeq2} to obtain $2^{2/3}g h_{(3)}^\alpha=-{\ast d\chi^\alpha_{(1)}}-(g/2)\epsilon_{\alpha\beta\gamma}{\ast F^{\beta\gamma}_{(2)}}$. Then the two linearised equations \eqref{redeq1},\eqref{redeq4}
can be combined into the form
\begin{align}
d{\ast d\chi^\alpha_{(1)}}+gd\,{\ast (\epsilon_{\alpha\beta\gamma}F^{\beta\gamma}_{(2)})}=0\,,\qquad
d{\ast d\chi^\alpha_{(1)}}=-2^{5/3}g^2{\ast\chi_{(1)}^\alpha}\,,
\end{align}
corresponding to a triplet of massless vectors, dual to conserved currents with $\Delta=3$, and
a triplet of massive vector operators with $\Delta=5$.

These operators can be arranged into multiplets of $SU(2,2|2)$. It is helpful to first identify the operators that
survive the truncation to Romans' theory, as discussed in section \ref{romans}. These consist
of the stress tensor, with $\Delta=4$, $SU(2)\times U(1)$ conserved currents with $\Delta=3$, the two two-forms
associated with operators with $\Delta =3$ and the scalar operator (coming from the $\phi,\lambda$ sector) with $\Delta=2$.
These form the bosonic operators of the superconformal supermultiplet of $SU(2,2|2)$ that contains
the stress tensor;  
this multiplet is denoted by $A_2\bar A_{\bar 2}[0;0]^{(0;0)~}_{\Delta=2}$ in (5.95) of 
\cite{Cordova:2016emh}.

The remaining operators are a scalar (coming from the $\phi,\lambda$ sector) with $\Delta=6$, 
five scalars (coming from $\mathcal{T}$) with $\Delta=4$, six scalars (coming from $\db^{a\alpha}$) with
$\Delta=5$ and a triplet of vector operators with $\Delta=5$.
These form the bosonic operators of a supermultiplet of $SU(2,2|2)$ that is denoted,
in the notation of section 4.6 of \cite{Cordova:2016emh}, as
$B_1\bar B_1$ with superconformal chiral primary $[0;0]^{(4;0)}_4$ (associated with the five scalars
with $\Delta=4$.)

We conclude this subsection by proving that there are no further maximally supersymmetric $AdS_5$ vacua.
In fact, given the gauge group is $SO(2)\times SE(3)$, the results of  \cite{Louis:2015dca,Bobev:2018sgr} imply that for $l=-1$ the
above vacuum is necessarily unique. For $l=0$ and $l=+1$, we need to analyse the conditions for
maximal supersymmetry as presented in \cite{Bobev:2018sgr}. Taking into account that \cite{Bobev:2018sgr} worked in a basis in which
$\eta$ was diagonal 
we first define
\begin{align}
\hat{f}^{ABC}&=f^{MNP}(\mathcal{U}\cdot \mathcal{V})^A_{\phantom{m}M}(\mathcal{U}\cdot \mathcal{V})^B_{\phantom{B}N}(\mathcal{U}\cdot \mathcal{V})^C_{\phantom{C}P}, \nn
\hat{\xi}^{AB}&=\xi^{MN}(\mathcal{U}\cdot \mathcal{V})^{A}_{\phantom{A}M}(\mathcal{U}\cdot \mathcal{V})^{B}_{\phantom{B}N}\,,
\end{align}
where the matrix $\mathcal{U}$ was defined in \eqref{usimmat}.
Decomposing the $A,B,C$ indices in a 5+3 split via e.g.
 $A=\{m,\hat a\}$ with
 $m\in \{1,\dots,5\}$ and $\hat a\in\{6,7,8\}$,
the necessary and sufficient conditions for supersymmetry are given by $\xi^{M}=0$
and in addition
\begin{align}\label{eq_conditions_N=4_vacuum}
\hat{\xi}^{[mn}\hat{\xi}^{pq]}=&\,0\,,\qquad \hat{\xi}^{m\hat a}=\,0\,,\nn
\hat{f}^{mn\hat a}=&\,0\,,\qquad
6\sqrt{2}\Sigma^3\hat{\xi}_{mn}=-\epsilon_{mnpqr}\hat{f}^{pqr}\,,
\end{align}
with $\hat{\xi}^{mn}$ and $\hat{f}^{mnp}$ not identically zero.
Given the embedding tensor coefficients in \eqref{embeddingoef} and the coset representative in \eqref{cosettext}
a calculation reveals that the conditions are indeed satisfied when $l=-1$ for the above maximally supersymmetric vacuum and furthermore, they cannot be satisfied when $l=0,+1$.

\subsection{Non-supersymmetric $AdS_5$ vacua}\label{nonsusy}
When $l=+1$ there are additional non-supersymmetric $AdS_5$ solutions.
The first was first found in \cite{Gauntlett:2002rv} and has
\begin{equation}
e^{6\phi}=\tfrac{1}{3}(215+59\sqrt{13})^{1/5}\,,\qquad
e^{10\lambda}=3+\sqrt{13}\,,
\end{equation}
with all other fields trivial, and the $AdS_{5}$ radius squared $L^2$ is given by
\begin{equation}
{g}^2 L^2=\tfrac{4}{3^{5/3}}(-35+13\sqrt{13})^{1/3}\,.
\end{equation}
It has already been shown in \cite{Gauntlett:2002rv} that the linearised perturbations in the $\phi,\lambda$ sector give rise to modes that violate the BF bound, and hence this solution 
is unstable.

The second solution, which is new, is found by numerically solving the equations of motion. It is a solution that lies
within the $SO(2)_D$ truncation \eqref{so2dsec} and again has $l=+1$ with
\begin{align}\label{phi,lambda,w_stationary_points}
\phi&\sim 0.00721714\,,\quad \lambda\sim 0.246758\,,\quad w\sim -0.107101\,,\nn
z^a z^a&\sim0.262789\,,\qquad {g}^2L^2\approx1.26882\,.
\end{align}
Since $z^a$ is non-zero, the solution spontaneously breaks the anti-diagonal $SO(2)$ gauge group (see \eqref{newcdv}).
By examining the linearised scalar perturbations of $\phi,\lambda,w, z^a$ within the $SO(2)_D$ truncation, we find five modes with
mass squared, $m^2$, given by
\begin{align}
m^2L^2\sim 30.4342\,,\quad 22.7531\,, \quad 9.44854\,,\quad -6.92312\,,
\end{align}
as well as zero (associated with the phase of $z^a$). In particular there is a mode which violates the BF bound $m^2L^2\ge -4$ and hence
this solution is also unstable.

\subsection{Supersymmetric $AdS_3$ and $AdS_2$ solutions}
There are a number of interesting solutions of Romans' theory that can be uplifted
to $D=11$ using the consistent truncation discussed in this paper. In fact these $D=11$ solutions
were already discussed in \cite{Gauntlett:2007sm}, so we shall be brief. From a dual field theory point of view,
the $D=11$ solutions describe RG flows of the $N=2$ SCFT in $d=4$ that is associated with M5-branes wrapping a two-dimensional hyperbolic space\footnote{As already discussed, we can also take discrete quotients of the $H^2$. We can similarly take quotients
of the $H^3,H^2, S^2$ and $\mathbb{R}^2$ factors that appear in the discussion below.} embedded in a Calabi-Yau two-fold, $H^2 \subset CY_2$.

We begin with the supersymmetric black hole solution, numerically constructed in \cite{Nieder:2000kc}, that flows from the supersymmetric $AdS_5$ vacuum in the UV to a supersymmetric $AdS_2\times H^3$ solution
in the IR.
The uplifted $D=11$ solution \cite{Gauntlett:2007sm} describes the RG flow
of the $N=2$, $d=4$ SCFT after being placed on $H^3$ with a topological twist that preserves
2 of the 8 Poincar\'e supersymmetries. In the far IR one obtains a supersymmetric conformal quantum mechanics
dual to the $AdS_2\times H^3\times H^2\times S^4$ solution (warped and fibred).
This $D=11$ $AdS_2$ solution is the one found in \cite{Gauntlett:2001jj} associated with M5-branes wrapping $(H^2\subset CY_2)\times (H^3\subset CY_3)$.

There is also supersymmetric black string solution of Romans theory, numerically constructed in \cite{Maldacena:2000mw}, that flows from 
the supersymmetric $AdS_5$ vacuum in the UV to an $AdS_3\times H^2$ solution in the IR.
The uplifted $D=11$ solution \cite{Gauntlett:2007sm} describes the RG flow
of the $N=2$, $d=4$ SCFT after being placed on $H^2$ with a topological twist that preserves, from a $d=2$ point of view,
$(2,2)$ of the 8 Poincar\'e supersymmetries. In the far IR one obtains a $d=2$, $(2,2)$ SCFT
dual to the $AdS_3\times H^2\times H^2\times S^4$ solution (warped and fibred).
This $D=11$ $AdS_3$ solution is the one found in \cite{Gauntlett:2001jj} associated with M5-branes wrapping $(H^2\subset CY_2)\times (H^2\subset CY_2)$.

There is a different supersymmetric black string solution, which is also a solution of minimal
gauged supergravity \cite{Caldarelli:1998hg}, that flows from 
the supersymmetric $AdS_5$ vacuum in the UV to a different $AdS_3\times H^2$ solution in the IR.
The uplifted $D=11$ solution \cite{Gauntlett:2007sm} describes the RG flow
of the $N=2$, $d=4$ SCFT after being placed on $H^2$ with a topological twist that preserves, from a $d=2$ point of view,
$(0,2)$ of the 8 Poincar\'e supersymmetries. In the far IR one obtains a $d=2$, $(0,2)$ SCFT
dual to the $AdS_3\times H^2\times H^2\times S^4$ solution (warped and fibred).
This $D=11$ $AdS_3$ solution is the one found in \cite{Gauntlett:2000ng} associated with M5-branes wrapping $H^2\times H^2\subset CY_4$.

Finally, going back to Romans' theory there is a one parameter family of supersymmetric $AdS_3\times M_2$ solutions with $M_2 =H^2$, $\mathbb{R}^2$ or $S^2$, depending on the value of the parameter \cite{Romans:1985ps}. 
Generically, the $D=11$ solutions \cite{Gauntlett:2007sm} are dual to $d=2$ SCFTs with $(0,2)$ supersymmetry and for a specific value of the parameter includes the $AdS_3\times H^2$ solution of minimal supergravity discussed in the previous paragraph. For another specific value one obtains the $AdS_3\times H^2$ solution that is dual to 
$d=2$ SCFTs with $(2,2)$ supersymmetry, discussed above. 
Supersymmetric black string solutions, flowing from the supersymmetric $AdS_5$ vacuum in the UV to the $AdS_3\times M_2$ solution
in the IR imply that the $AdS_3\times M_2\times H^2\times S^4$ solutions are dual to the $N=2$, $d=4$ SCFT after being placed on $M_2$ with a suitable topological twist. Such black string solutions can be constructed numerically
for various values of the parameter \cite{privcom}, extending\footnote{A class of $AdS_3\times M_2$ solutions $D=5$ STU gauged supergravity theory, with three $U(1)$'s, were constructed in \cite{Benini:2012cz,Benini:2013cda}. These include the one parameter family of solutions to Romans theory \cite{Romans:1985ps} that we are discussing here: for example, in section 3.1 of
\cite{Benini:2013cda} one can set $a_1=a_2\equiv a$, thus setting two of the gauge fields to be equal, and $\phi_2=0$. The parameter $a$ can be related to the parameter $x$ in section 3.4 of \cite{Gauntlett:2007sm} via $a=-lx/(4x-1)$.} 
the solutions constructed in \cite{Benini:2013cda}, which suggests that they probably exist for all
values of the parameter.

\section{Final comments}\label{secfincom}

The focus of this paper has been to construct a new consistent KK truncation of $D=11$ supergravity
on $\Sigma_2\times S^4$ where $\Sigma_2=S^2,\mathbb{R}^2$ or $H^2$, or a quotient thereof.
We have shown the resulting $D=5$ theory is an $N=4$ gauged supergravity theory coupled to
three vector multiplets. We have shown that the only maximally supersymmetric $AdS_5$ solution
(i.e. preserving 16 supersymmetries) of the $N=4$, $D=5$ theory occurs for 
$\Sigma_2=H^2$ and uplifts to the $AdS_5\times H^2\times S^4$ solution of 
\cite{Maldacena:2000mw}, dual to $N=2$ SCFTs in $d=4$ 
(after taking a quotient to get a compact $H^2/\Gamma$). 
We have also explored the possibility of whether or not there are additional $AdS_5$ solutions; we have shown that the theory
admits two additional non-supersymmetric solutions which uplift to $AdS_5\times S^2\times S^4$ solutions of $D=11$, both
of which are unstable. It would be of interest to complete this exploration, using the approach of \cite{Comsa:2019rcz}, for example, 
and, more generally, investigate other types of solutions of  the $N=4$, $D=5$ gauged theory.

This work is a natural extension of the consistent KK truncation
of $D=11$ supergravity on $\Sigma_3\times S^4$ down to an $N=2$ gauged supergravity
in $D=4$, where $S^3$, $\mathbb{R}^3$ or $\Sigma_3=H^3$ (or a quotient thereof) that was presented in \cite{Donos:2010ax}. In that case the fibration of the $S^4$ over $\Sigma_3$ is associated with M5-branes wrapped 
on a special Lagrangian $\Sigma_3$ in Calabi-Yau three-fold. It is clear that for each of the
different cases of M5-branes wrapping different supersymmetric cycles $\Sigma_k$
studied in \cite{Gauntlett:2000ng,Gauntlett:2001jj}
there will be an associated consistent KK truncation on $\Sigma_k\times S^4$
and it would be interesting to work out the details.
It would also be interesting to examine our result, as well these generalisations,
using the perspective of generalised geometry along the lines discussed in, for example,
\cite{Lee:2014mla,Hohm:2014qga,Malek:2017njj}. In particular this should provide a succinct way of determining the specific gauged supergravity theory that should arise.
In fact for the case we have considered in this paper, we have been informed that this will be discussed in \cite{Cassani:2019vcl}, finding the same gauging that we have here.

\subsection*{Acknowledgments}
We thank Nikolay Bobev, Thomas Fischbacher,
Victor Lekeu, Eoin O'Colgain, Hagen Triendl and Dan Waldram for discussions. 
CAR would like to thank the Theoretical Physics Group at Imperial College and also the Crete Center for Theoretical Physics for hospitality.
KCMC is supported by an Imperial College President's PhD Scholarship.
JPG is supported by the European Research Council under the European Union's Seventh Framework Programme (FP7/2007-2013), ERC Grant agreement ADG 339140. JPG is also supported by STFC grant ST/P000762/1, EPSRC grant EP/K034456/1, as a KIAS Scholar and as a Visiting Fellow at the Perimeter Institute. 

\appendix

\section{Equations of motion of $D=7$ gauged supergravity}\label{seca1}
The equations of motion for $D=7$ gauged supergravity arising from \eqref{d7lag} are given by
\begin{align} \label{h4eq}
&DS_\3^i  = g T_{ij}\, {\ast S_\3^j} + \tfrac{1}{8}  \ep_{i
{j_1}\cdots {j_4}} F_\2^{{j_1} {j_2}}\wedge\, F_\2^{{j_3} {j_4}} \,, \nn
& {D\Big(T^{-1}_{ik} T^{-1}_{j l} {\ast F_\2^{ij}}\Big)} =-2g \,
T^{-1}_{i[k} {*DT_{ l] i}}
-\tfrac{1}{2} \,
\ep_{i_1 i_2 i_3 k  l}\, F_\2^{i_1 i_2}\w T_{i_3j}\, {\ast S_\3^j} - S_\3^k\wedge S_\3^ l\,,\nn
& D\Big(\, T^{-1}_{ik} {*D(T_{kj}})\Big)
=2g^2 (2 T_{ik}\, T_{kj} -
T_{kk}\,
T_{ij})\mathrm{vol}_{7} + T^{-1}_{im}\, T^{-1}_{k l}\,
{\ast F_\2^{m l}}\wedge F_\2^{kj}
+T_{jk}\, {*S_\3^k} \wedge S_\3^i
  \nn
&  \quad   -\tfrac{1}{5} \delta_{ij}
\Big[2g^2\Big(2T_{ik} T_{ik} -  (T_{ii})^2 \Big) \mathrm{vol}_{7}  +
T^{-1}_{nm} T^{-1}_{k l}\, {\ast F_\2^{m l}} \wedge F_\2^{kn} +
 T_{k l } \, {*S_\3^k} \wedge S_\3^ l \Big]\,, 
 \end{align}
 and
 \begin{align}
& \label{7dEinstein}
R_{\mu\nu}=\tfrac{1}{4}T^{-1}_{ij}T_{kl}^{-1}D_\mu T_{jk} D_\nu T_{li}
+\tfrac{1}{4}T^{-1}_{ik}T^{-1}_{jl}F^{ij}_{\mu\rho}F^{kl\rho}_\nu
+\tfrac{1}{4}T_{ij}S^i_{\mu\rho_1\rho_2}S^{j\rho_1\rho_2}_\nu+\tfrac{1}{10}g_{\mu\nu}X\,,
\end{align}
where
\begin{align}
X&= -\tfrac{1}{4}T^{-1}_{ik}T^{-1}_{jl}F^{ij}_{\rho_1\rho_2}F^{kl\rho_1\rho_2}
-\tfrac{1}{3}T_{ij}S^i_{\rho_1\rho_2\rho_3}S^{j\rho_1\rho_2\rho_3}+2V\,.
\end{align}
We note that a typo in \cite{Cvetic:2000ah} has been fixed in the last equation of \eqref{h4eq}.

\section{Consistency of the truncation}\label{appb}

We substitute the ansatz for the $D=7$ fields given in \eqref{metans}-\eqref{3fans} into the
equations of motion for $D=7$ gauged supergravity given in \eqref{h4eq}-\eqref{7dEinstein}.
In carrying out the computations it is useful to note that for the scalars we have
\begin{align}
DT^{ab}&=-6e^{-6\lambda}d\lambda\delta^{ab}\,,\nn
DT^{a\alpha}&=g\left(e^{4\lambda}(\mathcal{T}\db^1)_\alpha-e^{-6\lambda}\db^1_\alpha\right)\bar{e}^a
-g\left(   e^{4\lambda}(\mathcal{T}\db^2)_{\alpha}-   e^{-6\lambda}\db^2_\alpha\right)\epsilon^{ab}\bar{e}^b\,,\nn
DT^{\alpha\beta}&=e^{4\lambda}\left(4d\lambda\mathcal{T}^{\alpha\beta}+D\mathcal{T}^{\alpha\beta}\right)\,,
\end{align}
where $D\mathcal{T}_{\alpha\beta}\equiv d\mathcal{T}_{\alpha\beta}+gA^{\alpha\gamma}_{(1)}\mathcal{T}_{\gamma\beta}+gA^{\beta\gamma}_{(1)}\mathcal{T}_{\alpha\gamma}$.
Furthermore, for the gauge fields we deduce
\begin{align}
F^{ab}_{(2)}&=g\left(l-\db^2\right)\,\bar{e}^a\wedge\bar{e}^b+\epsilon^{ab}F_{(2)}\,,\nn
F^{a\alpha}_{(2)}&=D\db^{1\alpha}\wedge\bar{e}^a-D\db^{2\alpha}\wedge \epsilon^{ab}\bar{e}^b\,,\nn
F^{\alpha\beta}_{(2)}&={F}^{\alpha\beta}_{(2)}
+2g(\epsilon^{ab}\db^{a\alpha}\db^{b\beta})\mathrm{vol}(\Sigma_2)\,,
\end{align}
where we have defined
\begin{align}
F_{(2)}&\equiv dA_{(1)}\,,\nn
{F}^{\alpha\beta}_{(2)}&\equiv dA^{\alpha\beta}_{(1)}+gA^{\alpha\gamma}_{(1)}\wedge A^{\gamma\beta}_{(1)}\,,\nn
D\db^{a\alpha}&\equiv d\db^{a\alpha}+gA^{\alpha\beta}_{(1)}\db^{a\beta}+gA_{(1)}\epsilon^{ab}\db^{b\alpha}\,.
\end{align}
Similarly, for the three-form we have 
\begin{align}
DS^a_{(3)}&=(D\K^1_{(2)}-g\db^{1\alpha} h^{\alpha}_{(3)})\wedge \bar{e}^a
-(D\K^2_{(2)}-g\db^{2\alpha} h^\alpha_{(3)})\wedge\epsilon^{ab}\bar{e}^b\,,\nn
DS^\alpha_{(3)}&=Dh^\alpha_{(3)}+(D\chi^{\alpha}_{(1)}
+2g\epsilon^{ab}\db^{a\alpha}\K^b_{(2)})\wedge \mathrm{vol}(\Sigma_2)\,,
\end{align}
where we have defined
\begin{align}
D\K^a_{(2)}&\equiv d\K^a_{(2)}+g\epsilon^{ab}A_{(1)}\wedge\K^b_{(2)}\,,\nn
Dh^\alpha_{(3)}&\equiv dh^\alpha_{(3)}+gA^{\alpha\beta}_{(1)}\wedge h^\beta_{(3)}\,,\nn
D\chi^{\alpha}_{(1)}&\equiv d\chi^\alpha_{(1)}+gA^{\alpha\beta}_{(1)}\wedge \chi^{\beta}_{(1)}\,.
\end{align}
Finally, for the metric sector, we use the orthonormal frame $e^m=e^{-2\phi}\bar e^m$, $m=1,...,5$ and 
$e^a=e^{3\phi}\bar e^a$, $a=1,2$ and find that the $D=7$ Ricci tensor has components
\begin{align}
R_{mn}&=e^{4\phi}\left( R^{(5)}_{mn}+2\nabla^2\phi \eta_{mn}-30\nabla_m\phi\nabla_n\phi\right)\,,\nn
R_{am}&=0\,,\nn
R_{ab}&=e^{4\phi}\left(-3\nabla^2\phi+lg^2e^{-10\phi}\right)\delta_{ab}\,,
\end{align}
where $ R^{(5)}_{mn}$ is the Ricci tensor for the $D=5$ metric $ds^2_5=\bar e^m\bar e^m$ in
\eqref{metans} and we used $R^{(2)}_{ab}=lg^2\delta_{ab}$, where $R^{(2)}_{ab}$ is
the Ricci tensor for $ds^2(\Sigma_2)=\bar e^a\bar e^a$.

\subsection{$D=5$ Equations of motion}

The equations of motion for the three-form in \eqref{h4eq} give rise to
\begin{align}\label{redeq1}
D\K_{(2)}^a-g\db^{a\alpha}h^{\alpha}_{(3)}&=-ge^{-6\lambda-2\phi}\epsilon^{ab}{\ast \K^b_{(2)}}
+\tfrac{1}{2}\epsilon_{\alpha\beta\gamma}\epsilon^{ab}D\db^{b\alpha}\wedge{F}^{\beta\gamma}_{(2)}\,,\nn
Dh^{\alpha}_{(3)}&=ge^{4\lambda-12\phi}{\ast (\mathcal{T}\chi_{(1)})^{\alpha}}+\tfrac{1}{2}\epsilon_{\alpha\beta\gamma}{F}^{\beta\gamma}_{(2)}\wedge F_{(2)}\,,
\end{align}
as well as
\begin{align}
\label{redeq2}
&D\chi^{\alpha}_{(1)}+2g\epsilon^{ab}\db^{a\alpha}\K^b_{(2)}=ge^{4\lambda+8\phi}{\ast (\mathcal{T}h_{(3)}})^{\alpha}\,,\nn
&\qquad\qquad\qquad+\epsilon_{\alpha\beta\gamma}\left(D\db^{a\beta}\wedge D\db^{a\gamma}
+\tfrac{1}{2}g(l-\db^2){F}^{\beta\gamma}_{(2)}+g\epsilon^{ab}\db^{a\beta}\db^{b\gamma}F_{(2)}\right)\,.
\end{align}
It is helpful to note that when $g\ne 0$ these imply
\begin{align}
D(e^{-6\lambda-2\phi}{\ast \K^a_{(2)}})=&-F_{(2)}\wedge \K^a_{(2)}-\epsilon^{ab}D\db^{b\alpha}\wedge h^{\alpha}_{(3)}
-g e^{4\lambda-12\phi}\epsilon^{ab}\db^{b\alpha}{\ast (\mathcal{T}\chi_{(1)}})^{\alpha}\,,\nn
D(e^{4\lambda-12\phi}{\ast (\mathcal{T}\chi_{(1)}})^{\alpha})=&\,{F}^{\alpha\beta}_{(2)}\wedge h^{\beta}_{(3)}\,,
\end{align}
and also
\begin{align}
&D(e^{4\lambda+8\phi}{\ast (\mathcal{T}h_{(3)}})^{\alpha})={F}^{\alpha\beta}_{(2)}\wedge \chi^\beta_{(1)}
+2\epsilon^{ab}D\db^{a\alpha} \wedge \K^b_{(2)}
+2g\epsilon^{ab}\db^{a\alpha}\db^{b\beta}h^\beta_{(3)}\nn
&\qquad\qquad\qquad\qquad\quad+2ge^{-6\lambda-2\phi}\db^{a\alpha} {\ast \K^a_{(2)}}\,,
\end{align}
where we used $\tfrac{1}{2}\epsilon_{\alpha\beta\gamma}{F}^{\alpha \rho}_{(2)} \wedge {F}^{\beta\gamma}_{(2)}=0$.

We next consider the gauge field equations of motion in  \eqref{h4eq}. When the indices $(k,l)=(a,b)$ and 
$(k,l)=(\alpha,\beta)$ we find
\begin{align}\label{redeq3}
&d(e^{12\lambda+4\phi}{\ast{}F_{(2)}})-2ge^{-6\phi+2\lambda}\epsilon^{ab}(\mathcal{T}^{-1}\db)^{a\alpha}{\ast{} D\db^{b\alpha}}
+\tfrac{1}{2}e^{4\lambda+8\phi}\epsilon_{\alpha\beta\gamma}{F}^{\alpha\beta}_{(2)}\wedge{\ast{} (\mathcal{T}h_{(3)}})^{\gamma}\nn
&\qquad\qquad+ge^{4\lambda-12\phi}\epsilon_{\alpha\beta\gamma}
(\epsilon^{ab}\db^{a\alpha}\db^{b\beta}){\ast{}(\mathcal{T}\chi_{(1)}})^{\gamma}
+\K^a_{(2)}\wedge \K^a_{(2)}=0\,,
\end{align}
and 
\begin{align}\label{redeq4}
&D(\mathcal{T}^{-1}_{\gamma[\alpha}\mathcal{T}^{-1}_{\beta]\rho}e^{4\phi-8\lambda}{\ast {F}^{\gamma\rho}_{(2)}})
-4ge^{2\lambda-6\phi}\db^{a[\alpha}(\mathcal{T}^{-1})^{\beta]\gamma}{\ast  D\db^{a\gamma}}
+2g\mathcal{T}^{-1}_{\gamma[\alpha}{\ast  D\mathcal{T}_{\beta]\gamma}}\nn
&
+\epsilon_{\alpha\beta\gamma}\bigg[g(l-\db^2)e^{4\lambda-12\phi}{\ast (\mathcal{T}\chi_{(1)}})^{\gamma}
+e^{4\lambda+8\phi}F_{(2)}\wedge {\ast  (\mathcal{T}h_{(3)}})^\gamma
-2e^{-6\lambda-2\phi}\epsilon^{ab}D\db^{a\gamma}\wedge{\ast  \K^b_{(2)}}\bigg]\nn
&+2h^{[\alpha}_{(3)}\wedge \chi^{\beta]}_{(1)}=0\,,
\end{align}
respectively. When the indices $(k,l)=(a,\alpha)$ we get 
\begin{align}\label{redeq5}
&D(e^{2\lambda-6\phi}\mathcal{T}^{-1}_{\alpha\beta}{\ast  D\db^{a\beta}})
-{g}^2\bigg[2e^{-8\lambda-16\phi}\epsilon^{ab}\epsilon^{cd}(\db^b \mathcal{T}^{-1}\db^d)
(\mathcal{T}^{-1} \db)^{c\alpha}\nn
&-e^{12\lambda-16\phi}(l-\db^2)\db^{a\alpha}
+e^{-10\phi}\left(e^{10\lambda}(\mathcal{T}\db)^{a\alpha}-2\db^{a\alpha}
+e^{-10\lambda}(\mathcal{T}^{-1}\db)^{a\alpha}\right)\bigg]\mathrm{vol}_5\nn
&+\epsilon_{\alpha\beta\gamma}\left(\tfrac{1}{2}e^{-6\lambda-2\phi}{F}^{\beta\gamma}_{(2)}
\wedge \epsilon^{ab}{\ast \K^b_{(2)}}
-e^{4\lambda-12\phi}{\ast (\mathcal{T}\chi_{(1)}})^\gamma\wedge D\db^{a\beta}\right)
+h^{\alpha}_{(3)}\wedge \epsilon^{ab}\K^b_{(2)}=0\,.
\end{align}

Continuing, we now consider the equations of motion for the scalar fields in \eqref{h4eq}.
From the $(i,j)=(a,b)$ components, we obtain: 
\begin{align}\label{redeq7}
&d({\ast  d\lambda})-\tfrac{1}{10}e^{4\phi+12\lambda}{\ast  F_{(2)}\wedge F_{(2)}}-\tfrac{1}{30}e^{8\phi+4\lambda}{\ast  h^{\alpha}_{(3)}}\wedge (\mathcal{T} h_{(3)})^\alpha
-\tfrac{1}{30}e^{4\lambda-12\phi}{\ast  \chi^{\alpha}_{(1)}}\wedge (\mathcal{T}\chi_{(1)})^\alpha\nn
&-\tfrac{1}{30}e^{2\lambda-6\phi}\mathcal{T}^{-1}_{\alpha\beta}{\ast  D\db^{a\alpha}}\wedge D\db^{a\beta}-\tfrac{1}{30}e^{4\phi-8\lambda}\mathcal{T}^{-1}_{\alpha\beta}\mathcal{T}^{-1}_{\gamma\rho}{\ast  {F}^{\beta\rho}_{(2)}}\wedge {F}^{\gamma\alpha}_{(2)}\nn
&+\tfrac{1}{10}e^{-6\lambda-2\phi}{\ast  \K^a_{(2)}}\wedge \K^a_{(2)}
+{g}^2\Bigg[\tfrac{1}{6}e^{-10\phi}\Big(e^{-10\lambda}(\db\mathcal{T}^{-1}\db)-e^{10\lambda}(\db\mathcal{T}\db)\Big)\nn
&-\tfrac{1}{15}e^{-4\phi}\left(2e^{8\lambda}\mathrm{Tr}(\mathcal{T}^2)-e^{8\lambda}(\mathrm{Tr}\mathcal{T})^2+e^{-2\lambda}\mathrm{Tr}\mathcal{T}\right)\nn
&-\tfrac{1}{10}\Big(l-\db^2\Big)^2 e^{12\lambda-16\phi}
+\tfrac{2}{15}e^{-8\lambda-16\phi}
\epsilon^{ab}\epsilon^{cd}
(\db^a\mathcal{T}^{-1}\db^c)(\db^b\mathcal{T}^{-1}\db^d)\Bigg]\mathrm{vol}_5=0\,.
\end{align}
From the $(i,j)=(\alpha,\beta)$ components, we obtain 
\begin{align}\label{redeq8}
&D(\mathcal{T}^{-1}_{\alpha\gamma}{\ast  D\mathcal{T}_{\gamma\beta}})
+\tfrac{2}{3}e^{2\lambda-6\phi}\Big(3\mathcal{T}^{-1}_{\alpha\gamma}\delta_{\beta\rho}-\mathcal{T}^{-1}_{\gamma\rho}\delta_{\alpha\beta}\Big){\ast  D\db^{a\gamma}\wedge D\db^{a\rho}}\nn
&-\tfrac{1}{3}e^{-8\lambda+4\phi}\Big(3\mathcal{T}^{-1}_{\alpha\gamma}\mathcal{T}^{-1}_{\rho\eta}\delta_{\beta\xi}-\mathcal{T}^{-1}_{\xi\gamma}\mathcal{T}^{-1}_{\rho\eta}\delta_{\alpha\beta}\Big)\,{\ast  {F}^{\gamma\eta}_{(2)}}\wedge {F}^{\rho\xi}_{(2)}\nn
&-\tfrac{1}{3}e^{4\lambda+8\phi}\Big(3\mathcal{T}_{\beta\gamma}\delta_{\alpha\rho}-\mathcal{T}_{\gamma\rho}\delta_{\alpha\beta}\Big)\, {\ast h^{\gamma}_{(3)}}\wedge h^{\rho}_{(3)}-\tfrac{1}{3}e^{4\lambda-12\phi}\Big(3\mathcal{T}_{\beta\gamma}\delta_{\alpha\rho}-\mathcal{T}_{\gamma\rho}\delta_{\alpha\beta}\Big)\, {\ast \chi^{\gamma}_{(1)}}\wedge \chi^{\rho}_{(1)}\nn
&+{g}^2\Bigg\{\tfrac{2}{3}e^{-10\phi}\Big[3e^{-10\lambda}(\mathcal{T}^{-1}\db)^{a\alpha} \db^{a\beta}
-3e^{10\lambda}\db^{a\alpha} (\mathcal{T}\db)^{a\beta}
-e^{-10\lambda}(\db\mathcal{T}^{-1}\db)\delta_{\alpha\beta}+e^{10\lambda}(\db\mathcal{T}\db)\delta_{\alpha\beta}\Big]
\nn
&+\tfrac{2}{3}e^{-4\phi}\Big[2e^{8\lambda}\mathrm{Tr}(\mathcal{T}^2)\delta_{\alpha\beta}-e^{8\lambda}(\mathrm{Tr}\mathcal{T})^2\delta_{\alpha\beta}-2e^{-2\lambda}\mathrm{Tr}\mathcal{T}\delta_{\alpha\beta}\nn
&\qquad\qquad\qquad\qquad-6e^{8\lambda}(\mathcal{T}^2)_{\alpha\beta}+3e^{8\lambda}\mathrm{Tr}\mathcal{T}\mathcal{T}_{\alpha\beta}+6e^{-2\lambda}\mathcal{T}_{\alpha\beta}\Big]\nn
&-\tfrac{4}{3}e^{-8\lambda-16\phi}\Big[3\mathcal{T}^{-1}_{\alpha\gamma}\mathcal{T}^{-1}_{\rho\eta}\delta_{\beta\xi}-\mathcal{T}^{-1}_{\xi\gamma}\mathcal{T}^{-1}_{\rho\eta}\delta_{\alpha\beta}\Big]
(\epsilon^{ab}\db^{a\gamma}\db^{b\eta})(\epsilon^{cd}\db^{c\rho}\db^{d\xi})\Bigg\}\mathrm{vol}_{5}=0\,.
\end{align}
The equations of motion for the scalar fields with mixed components $(i,j)=(a, \alpha)$ are trivially satisfied.

Finally, we consider the reduction of the Einstein equations \eqref{7dEinstein}. 
From the $(a,b)$ components, we obtain
\begin{align}\label{redeq9}
&d({\ast  d\phi})-\tfrac{1}{30}e^{12\lambda+4\phi} {\ast  F_{(2)}}\wedge F_{(2)}
+\tfrac{1}{10}e^{2\lambda-6\phi}\mathcal{T}^{-1}_{\alpha\beta}{\ast  D\db^{a\alpha}}\wedge D\db^{a\beta}\nn
&-\tfrac{1}{60}e^{-8\lambda+4\phi}\mathcal{T}^{-1}_{\alpha\beta}\mathcal{T}^{-1}_{\gamma\rho}{\ast  {F}^{\alpha \gamma}_{(2)}}\wedge {F}^{\beta\rho}_{(2)}
+\tfrac{1}{30}e^{-6\lambda-2\phi}{\ast  \K^a_{(2)}}\wedge \K^a_{(2)}\nn
&+\tfrac{1}{10}e^{4\lambda-12\phi}{\ast  \chi^{\alpha}_{(1)}}\wedge (\mathcal{T}\chi_{(1)})^\alpha
-\tfrac{1}{15}e^{4\lambda+8\phi}{\ast  h^{\alpha}_{(3)}}\wedge (\mathcal{T}h_{(3)})^\alpha\nn
&+{g}^2 \Bigg\{\tfrac{1}{6}e^{-10\phi}\left(e^{10\lambda}(\db\mathcal{T}\db)
-2(l+\db^2)+e^{-10\lambda}(\db\mathcal{T}^{-1}\db) \right)
 +\tfrac{2}{15}e^{12\lambda-16\phi}\left(l-\db^2\right)^2\nn
&
\qquad+\tfrac{1}{30}e^{-4\phi}\left(2e^{8\lambda}\mathrm{Tr}(\mathcal{T}^2)-e^{8\lambda}\left(\mathrm{Tr}\mathcal{T}\right)^2-4e^{-2\lambda}\mathrm{Tr}\mathcal{T}\right)\nn
&\qquad
+\tfrac{4}{15}e^{-8\lambda-16\phi}\epsilon^{ab}\epsilon^{cd}(\db^a\mathcal{T}^{-1}\db^c  )( \db^b\mathcal{T}^{-1} \db^d )
\Bigg\}\mathrm{vol}_5=0\,.
\end{align}
From the ($m,n$) components we find that
the $D=5$ Ricci tensor must satisfy
\begin{align}\label{redeq10}
{R}^{(5)}_{mn}=&\,30{\nabla_m\phi}{\nabla_n\phi}+30{\nabla_m\lambda}{\nabla_n\lambda}+\tfrac{1}{4}\mathcal{T}^{-1}_{\alpha\beta}\mathcal{T}^{-1}_{\gamma\rho}D_m\mathcal{T}_{\beta\gamma}D_n\mathcal{T}_{\rho\alpha}\nn
&+\tfrac{1}{2}e^{12\lambda+4\phi}\left((F_{(2)})_{ml}(F_{(2)})^{\phantom{n}l}_{n}-\tfrac{1}{6}g_{mn}(F_{(2)})_{ls}(F_{(2)})^{ls}\right)\nn
&+e^{-6\lambda-2\phi}\left((\K_{(2)}^a)_{ml}(\K_{(2)}^a)^{\phantom{n}l}_{n}-\tfrac{1}{6}g_{mn}(\K_{(2)}^a)_{ls}(\K_{(2)}^a)^{ls}\right)\nn
&+\tfrac{1}{4}e^{-8\lambda+4\phi}\mathcal{T}^{-1}_{\alpha\beta}\mathcal{T}^{-1}_{\gamma\rho}\left(({F}^{\alpha\gamma}_{(2)})_{ml}({F}^{\beta\rho}_{(2)})_{n}^{\phantom{n}l}-\tfrac{1}{6}g_{mn}({F}^{\alpha\gamma}_{(2)})_{ls}({F}^{\beta\rho}_{(2)})^{ls}\right)\nn
&+e^{2\lambda-6\phi}\mathcal{T}^{-1}_{\alpha\beta}\left(D_m\db^{a\alpha} D_n\db^{a\beta}\right)+\tfrac{1}{2}e^{4\lambda-12\phi}(\chi^{\alpha}_{(1)})_m(\mathcal{T}\chi_{(1)})^\alpha_n\nn
&+\tfrac{1}{4}e^{4\lambda+8\phi}\mathcal{T}_{\alpha\beta}\left((h^{\alpha}_{(3)})_{mls}(h^{\beta}_{(3)})_{n}^{\phantom{n}ls}-\tfrac{2}{9}g_{mn}(h^{\alpha}_{(3)})_{lst}(h^{\beta}_{(3)})^{lst}\right)\nn
&+{g}^2g_{mn}\Bigg\{\tfrac{1}{6}e^{-4\phi}\left(2e^{8\lambda}\mathrm{Tr}(\mathcal{T}^2)-e^{8\lambda}(\mathrm{Tr}\mathcal{T})^2-4e^{-2\lambda}\mathrm{Tr}\mathcal{T}\right)\nn
&+\tfrac{1}{6}e^{12\lambda-16\phi}(l-\db^2)^2
+\tfrac{1}{3}e^{-8\lambda-16\phi}\epsilon^{ab}\epsilon^{cd}(\db^a\mathcal{T}^{-1}\db^c  )( \db^b\mathcal{T}^{-1} \db^d )\nn
&-\tfrac{1}{3}e^{-10\phi}\left(2(l+\db^2)
-e^{10\lambda}(\db\mathcal{T}\db)
-e^{-10\lambda}(\db\mathcal{T}^{-1}\db)\right)\Bigg\}\,.
\end{align}
The mixed ($ma$) components are trivially satisfied.

\subsection{Subtruncation to Romans' theory}
If we consider the subtruncation considered in section \ref{romans} then we find that
the $D=5$ equations of motion given in \eqref{redeq1}-\eqref{redeq2} and \eqref{redeq3}-\eqref{redeq10}
boil down to
\begin{align}\label{romanseq1}
D\mathcal{C}_{(2)}=&\,ige^{-20\phi}{\ast \mathcal{C}_{(2)}}\,,\nn
d\left(e^{40\phi}{\ast F_{(2)}}\right)=&-\tfrac{1}{2}{F}^{\alpha\beta}_{(2)}\wedge{F}^{\alpha\beta}_{(2)}-\overline{\mathcal{C}}_{(2)}\wedge \mathcal{C}_{(2)}\,,\nn
D\left(e^{-20\phi} {\ast {F}^{\alpha\beta}_{(2)}}\right)=&-{F}^{\alpha\beta}_{(2)}\wedge F_{(2)}\,,\nn
d{\ast d\phi}=&\,\tfrac{1}{30}e^{40\phi}{\ast F_{(2)}}\wedge F_{(2)}-\tfrac{1}{30}e^{-20\phi}{\ast \overline{\mathcal{C}}_{(2)}}\wedge\mathcal{C}_{(2)}\,,\nn
&-\tfrac{1}{60}e^{-20\phi}{\ast {F}^{\alpha\beta}_{(2)}}\wedge{F}^{\alpha\beta}_{(2)}-\tfrac{1}{30}{g}^2\left(e^{20\phi}-2e^{-10\phi}\right)\text{vol}_{5}\,,
\end{align}
and
\begin{align}\label{romanseq2}
{R}_{mn}&=300{\nabla_m\phi}{\nabla_n\phi}+\tfrac{1}{2}e^{40\phi}\left((F_{(2)})_{ml}(F_{(2)})^{\phantom{n}l}_{n}-\tfrac{1}{6}g_{mn}(F_{(2)})_{ls}(F_{(2)})^{ls}\right)\nn
&+\tfrac{1}{2}e^{-20\phi}\left(({F}^{\alpha\beta}_{(2)})_{ml}({F}^{\alpha\beta}_{(2)})_{n}{}^{l}-\tfrac{1}{6}g_{mn}({F}^{\alpha\beta}_{(2)})_{ls}({F}^{\alpha\beta}_{(2)})^{ls}\right)
-\tfrac{1}{3}{g}^2g_{mn}\left(4e^{-10\phi}+e^{20\phi}\right)\nn
&+e^{-20\phi}\left((\mathcal{C}_{(2)})_{(m|l|}(\overline{\mathcal{C}}_{(2)})^{\phantom{n)}l}_{n)}-\tfrac{1}{6}g_{mn}(\mathcal{C}_{(2)})_{ls}(\overline{\mathcal{C}}_{(2)})^{ls}\right)\,.
\end{align}
In these expressions we have $\mathcal{C}_{(2)}=\K^1_{(2)}+i\K^2_{(2)}$ with 
$D\mathcal{C}_{(2)}=d\mathcal{C}_{(2)}-igA_{(1)}\wedge \mathcal{C}_{(2)}$.
These equations of motion can be derived from the Lagrangian given in \eqref{roamnstextlag}.

\section{Matching with $N=4$ supergravity}\label{appc}

We present a few formulae which are helpful in explicitly matching the reduced $D=5$ theory of section \ref{sec3} with those of $N=4$, $D=5$ gauged supergravity theory that was discussed in section \ref{sec4pt1}.

We begin by clarifying the parametrisation of the $SL(3)/SO(3)$ coset that we used in \eqref{sl3co3coset}
The generators for the Lie algebra of $SL(3)$ are given by
\begin{align}
&\boldsymbol{{{h}_{1}}}=
\begin{pmatrix}
1 & 0 &0\\
0 & -1 &0 \\
0 & 0 &0
\end{pmatrix}\,,
\quad
\boldsymbol{{{h}_{2}}}=
\begin{pmatrix}
0 & 0 &0\\
0 & 1 &0 \\
0 & 0 &-1
\end{pmatrix}\,,
\nn
&\boldsymbol{{{e}_{1}}}=
\begin{pmatrix}
0 & 1 &0\\
0 & 0 &0 \\
0 & 0 &0
\end{pmatrix}\,,
\quad
\boldsymbol{{{e}_{2}}}=
\begin{pmatrix}
0 & 0 &0\\
0 & 0 &1 \\
0 & 0 &0
\end{pmatrix}\,,
\quad
\boldsymbol{{{e}_{3}}}=
\begin{pmatrix}
0 & 0 &1\\
0 & 0 &0 \\
0 & 0 &0
\end{pmatrix}\,,
\nn
&\boldsymbol{{{f}_{1}}}=
\begin{pmatrix}
0 & 0 &0\\
1 & 0 &0 \\
0 & 0 &0
\end{pmatrix}\,,
\quad
\boldsymbol{{{f}_{2}}}=
\begin{pmatrix}
0 & 0 &0\\
0 & 0 &0 \\
0 & 1 &0
\end{pmatrix}\,,
\quad
\boldsymbol{{{f}_{3}}}=
\begin{pmatrix}
0 & 0 &0\\
0 & 0 &0 \\
1 & 0 &0
\end{pmatrix}\,.
\end{align}
The coset element can then be represented in an upper triangular gauge via
\begin{align}
V&=e^{\varphi_1\boldsymbol{{{h}_{1}}}+\varphi_2\boldsymbol{{{h}_{2}}}}e^{a_1\boldsymbol{{{e}_{1}}}}e^{a_2\boldsymbol{{{e}_{2}}}}e^{a_3\boldsymbol{{{e}_{3}}}}\,,\nn
&=\left(
\begin{array}{ccc}
 e^{\varphi _1} & e^{\varphi _1} a_1 & e^{\varphi _1} \left(a_1 a_2+a_3\right) \\
 0 & e^{\varphi _2-\varphi _1} & e^{\varphi _2-\varphi _1} a_2 \\
 0 & 0 & e^{-\varphi _2} \\
\end{array}
\right)\,.
\end{align}

Next, turning to the $SO(5,3)/(SO(5)\times SO(3))$ coset element $\mathcal{V}$, given in \eqref{cosettext},
we find that the Maurer-Cartan one-form, which takes values in the solvable Lie algebra, has the form
\begin{align}\label{mcformone}
&d\mathcal{V}\cdot \mathcal{V}^{-1}=\nn
&\tfrac{1}{\sqrt{2}}d\varphi_1\mathscr{H}^1+\tfrac{1}{\sqrt{2}}d\varphi_2\mathscr{H}^2+\tfrac{1}{\sqrt{2}}d\varphi_3\mathscr{H}^3+e^{2\varphi_1-\varphi_2}da_1\mathcal{E}^1+e^{2\varphi_2-\varphi_1}da_2\mathcal{E}^2+e^{\varphi_1+\varphi_2}(da_3+a_1da_2)\mathcal{E}^3\nn
&+e^{-\varphi_2-2\varphi_3}X^3T^4
+e^{-\varphi_1+\varphi_2-2\varphi_3}( -X^2-a_2X^3)T^5
+e^{\varphi_1-2\varphi_3}(X^1+a_1X^2+(a_3+a_1a_2)X^3)T^6\nn
&+\sqrt{2}e^{-\varphi_1-\varphi_3}d\psi^{11}T^7
+\sqrt{2}e^{\varphi_1-\varphi_2-\varphi_3}(d\psi^{12}-a_1d\psi^{11})T^8
+\sqrt{2}e^{\varphi_2-\varphi_3}(d\psi^{13}-a_3d\psi^{11}-a_2d\psi^{12})T^9\nn
&+\sqrt{2}e^{-\varphi_1-\varphi_3}d\psi^{21}T^{10}
+\sqrt{2}e^{\varphi_1-\varphi_2-\varphi_3}(d\psi^{22}-a_1d\psi^{21})T^{11}
+\sqrt{2}e^{\varphi_2-\varphi_3}(d\psi^{23}-a_3d\psi^{21}-a_2d\psi^{22})T^{12}\,,
\end{align}
where
\begin{align}
X^\alpha\equiv d\xi^\alpha+\epsilon_{\alpha\beta\gamma}\psi^{a\beta}d\psi^{a\gamma}\,.
\end{align}
We can decompose the Maurer-Cartan one-form as 
\begin{align}
d\mathcal{V}\cdot\mathcal{V}^{-1}=\mathcal{P}^0+\mathcal{Q}^0\,,
\end{align}
where $\mathcal{Q}^0$ lies in the Lie algebra of $SO(5)\times SO(3)$ (the 
antisymmetric part of the one-form) and $\mathcal{P}^0$ lies in the complement
(the symmetric part of the one-form). We can then calculate
\begin{align}
\frac{1}{8}{\ast d \mathcal{M}_{MN}}\wedge d  \mathcal{M}^{MN}&=-\frac{1}{2}\text{Tr}(\ast \mathcal{P}^0\wedge
\mathcal{P}^0)\,,\nn
&=
-\frac{1}{4}\text{Tr}({\ast [d\mathcal{V}\cdot\mathcal{V}^{-1}}]\wedge [d\mathcal{V}\cdot\mathcal{V}^{-1}+(d\mathcal{V}\cdot\mathcal{V}^{-1})^T])\,,
\end{align}
and we obtain the kinetic terms for the scalars as in \eqref{lscalred}, without yet incorporating the 
gauging. To incorporate the latter we use the covariant derivative given in \eqref{eq_cov_derivative2}
which we write as $D=d+g\mathfrak{A}$ with
\begin{align}
\mathfrak{A}\equiv A_{\mu}\mathfrak{g}_0+A^1_{\mu}\mathfrak{g}_1+A^2_{\mu}\mathfrak{g}_2+A^3_{\mu}\mathfrak{g}_3+\mathscr{A}^1_{\mu}\mathfrak{g}_4+\mathscr{A}^1_{\mu}\mathfrak{g}_5+\mathscr{A}^3_{\mu}\mathfrak{g}_6\,.
\end{align}
We can then decompose $D\mathcal{V}\cdot\mathcal{V}^{-1}=\mathcal{P}+\mathcal{Q}$ as above.
In particular we have 
$\mathcal{P}=\mathcal{P}^0+g(\mathcal{V}\cdot \mathfrak{A}\cdot \mathcal{V}^{-1})_{SO(5,3)/(SO(5)\times SO(3))}$, where the last term is in the Lie algebra
complementary to that of $SO(5)\times SO(3)$.
We find that the gauged scalar kinetic terms in \eqref{lscalred} are obtained precisely after
calculating $-\frac{1}{2}\text{Tr}(\ast \mathcal{P}\wedge
\mathcal{P})$.

We can write the matrix $\mathcal{M}_{MN}$ in \eqref{emmmat} in the explicit form
\begin{align}\label{emmexplicit}
\mathcal{M}_{MN}
=&\left(
\begin{array}{ccc}
e^{-2\varphi _3}\mathcal{T}^{-1}& e^{-2\varphi _3}\mathcal{T}^{-1}\cdot\mathcal{S}^T & e^{-2\varphi _3}\mathcal{T}^{-1}\cdot\mathcal{Y}  \\
 e^{-2\varphi _3}\mathcal{S}\cdot\mathcal{T}^{-1} & e^{-2\varphi _3}\mathcal{S}\cdot\mathcal{T}^{-1}\cdot \mathcal{S}^T+ \mathds{1}_{2\times2} & e^{-2\varphi _3}\mathcal{S}\cdot\mathcal{T}^{-1}\cdot\mathcal{Y}+\mathcal{S} \\
 e^{-2\varphi _3}\mathcal{Y}^T\cdot\mathcal{T}^{-1} & e^{-2\varphi _3}\mathcal{Y}^T\cdot\mathcal{T}^{-1}\cdot\mathcal{S}^T+\mathcal{S}^T & e^{-2\varphi _3}\mathcal{Y}^T\cdot\mathcal{T}^{-1}\cdot \mathcal{Y}+\mathcal{S}^T\cdot\mathcal{S}+e^{2\varphi _3}\mathcal{T}\\
\end{array}
\right)\,,
\end{align}
where 
\begin{align}
\mathcal{S}_a{}^\alpha&\equiv \sqrt{2}\psi^{a\alpha}\,,\nn
\mathcal{Y}_{\alpha\beta}&\equiv \epsilon_{\alpha\beta\gamma}\xi^\gamma+\tfrac{1}{2}\mathcal{S}^\alpha_a\mathcal{S}^\beta_a\,.
\end{align}
To calculate the $N=4$ scalar potential $\mathcal{L}^{pot}_{N=4}$, given in \eqref{n4scalpot}, with the embedding tensor given in \eqref{embeddingoef}, 
we find the following non-vanishing contributions
\begin{align}\label{eq_schon_potential_pieces}
&-\tfrac{1}{2}f_{MNP}f_{QRS}\Sigma^{-2}\left(\tfrac{1}{12}\mathcal{M}^{MQ}\mathcal{M}^{NR}\mathcal{M}^{PS}-\tfrac{1}{4}\mathcal{M}^{MQ}\eta^{NR}\eta^{PS}+\tfrac{1}{6}\eta^{MQ}\eta^{NR}\eta^{PS}\right)\nn
&\qquad\qquad=-\tfrac{1}{2}e^{12\lambda-16\phi}(l-\db^2)^2+\tfrac{1}{2}e^{-4\phi+8\lambda}[(\mathrm{Tr}\mathcal{T})^2-2\mathrm{Tr}(\mathcal{T}^2)]\nn
&\qquad\qquad\quad
-e^{-10\phi+10\lambda}(\db\mathcal{T}\db)\,,\nn
&-\frac{1}{8}\xi_{MN}\xi_{PQ}\Sigma^4\left(\mathcal{M}^{MP}\mathcal{M^{NQ}}-\eta^{MP}\eta^{NQ}\right)\nn
&\qquad\qquad=-e^{-10\phi-10\lambda}(\db\mathcal{T}^{-1}\db)-e^{-8\lambda-16\phi}\epsilon^{ab}\epsilon^{cd}(\db^{a}\mathcal{T}^{-1}\db^{c})
(\db^{b}\mathcal{T}^{-1}\db^{d})\,,
\end{align}
and
\begin{align}\label{mfiveexpl}
&-\frac{1}{3\sqrt{2}}f_{MNP}\xi_{QR}\Sigma\mathcal{M}^{MNPQR}=\,2le^{-10\phi}+2e^{-10\phi}\db^2+2e^{-2\lambda-4\phi}\text{Tr}{\mathcal{T}}\,,
\end{align}
where in the last expression we have utilised the definition \eqref{mfiveddefu}. Summing these contributions
we find that the $N=4$ scalar potential $\mathcal{L}^{pot}_{N=4}$ in \eqref{n4scalpot} precisely gives the scalar potential 
$\mathcal{L}^{pot}$ 
of the reduced theory, given in \eqref{ellpotnf}.

Turning now to the vectors, using the identification of the field strengths given in
\eqref{eq.vector_identification} as well as \eqref{emmexplicit}, the kinetic terms of the vectors of the $N=4$ theory, $\mathcal{L}^V_{N=4}$, given
in \eqref{n4lag4}, exactly reproduce  the kinetic terms of the vectors in the reduced theory, $\mathcal{L}^{V}$, given in 
\eqref{Eq. new vector ke completr}. We next compare the topological parts of the Lagrangians.
We find that the non-zero contributions to $\mathcal{L}^{T}_{N=4}$, given in 
\eqref{n4lag5}, are (up to a total derivative),
\begin{align}\label{eq_schon_topological_pieces}
&\qquad\qquad\qquad-\frac{1}{\sqrt{2}}gZ^{\mathcal{M}\mathcal{N}}\mathcal{B}_{\mathcal{M}}\wedge D\mathcal{B}_{\mathcal{N}}=\tfrac{1}{2g}{L}^1_{(2)}\wedge D{L}^2_{(2)}-\tfrac{1}{2g}{L}^2_{(2)}\wedge D{L}^1_{(2)}\,,\nn
&\,\,\,\qquad\qquad\frac{\sqrt{2}}{3}d_{\mathcal{M}\mathcal{N}\mathcal{P}}\mathcal{A}^{\mathcal{M}}\wedge d\mathcal{A}^{\mathcal{N}}\wedge d\mathcal{A}^{\mathcal{P}}=-d[\mathscr{A}^\alpha_{(1)}-lA^\alpha_{(1)}]\wedge dA^\alpha_{(1)}\wedge A_{(1)}\,,\nn
&\frac{1}{2\sqrt{2}}{g}d_{\mathcal{M}\mathcal{N}\mathcal{P}}X_{\mathcal{Q}\mathcal{R}}^{\phantom{\mathcal{Q}\mathcal{R}}\mathcal{M}}\mathcal{A}^{\mathcal{N}}\wedge \mathcal{A}^{\mathcal{Q}}\wedge \mathcal{A}^{\mathcal{R}}\wedge d\mathcal{A}^{\mathcal{P}}=\nn
&\qquad\qquad\qquad\qquad\qquad\qquad-\tfrac{1}{2}g\epsilon_{\alpha\beta\gamma}d[\mathscr{A}^\alpha_{(1)}-lA^\alpha_{(1)}]\wedge A^\gamma_{(1)}\wedge A^\beta_{(1)}\wedge A_{(1)}\nn
&\qquad\qquad\qquad\qquad\qquad\qquad-g\epsilon_{\alpha\beta\gamma}A^\gamma_{(1)}\wedge[\mathscr{A}^\beta_{(1)}-\tfrac{1}{2}lA^\beta_{(1)}]\wedge dA^\alpha_{(1)}\wedge A_{(1)}\,.
\end{align}
Combining these expressions we recover the topological Lagrangian $\mathcal{L}^{T}$ of the reduced theory given in
\eqref{eq:new_top_after_field_redef}.

\section{Matching the $SO(2)_D$ truncation with $N=2$ supergravity}\label{appd}

We begin by discussing the quaternionic K\"ahler manifold $SU(2,1)/S[U(2)\times U(1)]$ 
(see e.g. \cite{Strominger:1997eb,Lukas:1998tt}).
An element $U$ of $SU(2,1)$ obeys $U^\dagger\eta U = \eta$ where
we take $\eta$ to have signature $(-,-,+)$. A convenient choice of the generators, satisfying
$\eta T=T^\dagger\eta$ is given by
\begin{equation}
T_i=\{\lambda_1,\lambda_2,\lambda_3,\lambda_8,i\lambda_4,i\lambda_5,i\lambda_6,i\lambda_7\}\,,
\end{equation}
where $\lambda_i$ are the standard Gell-Mann matrices. To construct a convenient coset representative we utilise one non-compact Cartan generator, $\boldsymbol{{{h}}}$, along with three positive root generators, $(\boldsymbol{{{r}_{1}}},\boldsymbol{{{r}_{2}}},\boldsymbol{{{r}_{3}}})$, given by
\begin{align}
\boldsymbol{{{h}}}=
\begin{pmatrix}
1 & 0 &0\\
0 & 0 &0 \\
0 & 0 &-1
\end{pmatrix}\,,
\,\,
\boldsymbol{{{r}_{1}}}=
\begin{pmatrix}
0 & 0 &0\\
0 & 0 &1 \\
0 & 0 &0
\end{pmatrix}\,,
\,\,
\boldsymbol{{{r}_{2}}}=
\begin{pmatrix}
0 & 1 &0\\
0 & 0 &0 \\
0 & 0 &0
\end{pmatrix}\,,
\,\,
\boldsymbol{{{r}_{3}}}=
\begin{pmatrix}
0 & 0 &1\\
0 & 0 &0 \\
0 & 0 &0
\end{pmatrix}\,.
\end{align}
The coset representative is then defined as
\begin{align}
V=&\,e^{\varphi\boldsymbol{{{h}}}}e^{2z^1\boldsymbol{{{r}_{1}}}+2z^2\boldsymbol{{{r}_{2}}}+\xi\boldsymbol{{{r}_{3}}}}\,,
\end{align}
with an associated Maurer-Cartan one-form given by
\begin{align}
&dV\cdot {V}^{-1}=d\varphi\boldsymbol{{h}}+2e^{\varphi}(dz^1\boldsymbol{{{r}_{1}}}+dz^2\boldsymbol{{{r}_{2}}})+e^{2\varphi}\left(d\xi-2z^1dz^2+2z^2dz^1\right)\boldsymbol{{{r}_{3}}}\,.
\end{align}
We can then calculate
\begin{align}
&\text{Tr}\left[{\ast (d{V}\cdot {V}^{-1}})\wedge (d{V}\cdot{V}^{-1}+(d{V}\cdot {V}^{-1})^\dagger \right]
=g_{XY}{\ast d q^X}\wedge dq^Y\,,
\end{align}
where the quaternionic K\"ahler metric is given by
\begin{align}\label{quatmetap}
g_{XY}dq^Xdq^Y=\,4d\varphi^2+4e^{2\varphi}{dz^a} dz^a+e^{4\varphi}(d\xi-2\epsilon_{ab}z^adz^b)^2\,,
\end{align}
with $q^X=(\varphi,\xi,z^1,z^2)$.

In order to display the quaternionic K\"ahler structure, we can introduce the following vierbein
\begin{equation}
f^1=2d\varphi\,,\quad f^2=e^{2\varphi}(d\xi-2\epsilon_{ab}z^adz^b)\,,\quad f^3=2e^{\varphi}dz^1\,,\quad f^4=2e^{\varphi}dz^2\,,
\end{equation}
with associated spin connection, satisfying $df^A+\omega_{(1)}^A{}_B\wedge f^B=0$, given by
\begin{equation}
\omega_{(1)}=\frac{1}{2}\left[(2M_{21}+M_{34})f^2+(M_{31}+M_{24})f^3+(M_{41}+M_{32})f^4\right]\,,
\end{equation}
where $M_{mn}=E_{mn}-E_{nm}$ are the generators of $SO(4)\sim SU(2)\times Sp(2)$, with $E_{mn}$ a $4\times 4$ matrix with 1 in the $m,n$ position and zeroes elsewhere.
To proceed we explicitly extract the $SU(2)$ factor by defining the matrices
$I_i=-\eta^i$ and $\bar I_i=-\bar\eta^i$, where $\eta,\bar\eta$ are the 't Hooft symbols. Explicitly, we have
\begin{align}
{I}_1=M_{41}+M_{32}\,,\quad{I}_2=M_{42}+M_{13}\,,\quad {I}_3=M_{21}+M_{43}\,,\nn
\bar{I}_1=M_{14}+M_{32}\,,\quad\bar{I}_2=M_{24}+M_{13}\,,\quad \bar{I}_3=M_{21}+M_{34}\,,
\end{align}
which satisfy $[{I}_i,{I}_j]=2\epsilon_{ijk}{I}_k$, $[\bar{I}_i,\bar{I}_j]=2\epsilon_{ijk}\bar{I}_k$ and $[{I}_i,\bar{I}_j]=0$. The spin connection can then be written as
\begin{equation}
\omega_{(1)}=\tfrac{1}{4}{I}_3f^2-\tfrac{1}{2}I_2 f^3+\tfrac{1}{2}I_1 f^4+\tfrac{3}{4}\bar{I}_3f^2
\end{equation}
and we denote the $SU(2)$ component, generated by the ${I}_i$, as $\vec{\omega}=(\tfrac{1}{2}f^4,-\tfrac{1}{2}f^3,\tfrac{1}{4}f^2)$. The curvature 2-form for the metric is given by
\begin{align}
R_{(2)}=&\,\tfrac{1}{4}{I}_3 (f^{12}+f^{34})-\tfrac{1}{4}{I}_2(f^{13}-f^{24})+\tfrac{1}{4}{I}_1(f^{14}+f^{23})+\tfrac{3}{4}\bar{I}_3(f^{12}-f^{34})\,,
\end{align}
where $f^{ij}\equiv f^i\wedge f^j$, and as a result we identify the $SU(2)$ factor as
\begin{equation}
\vec{R}=\tfrac{1}{4}(f^{14}+f^{23},f^{24}-f^{13},f^{12}+f^{34})\,.
\end{equation}
It is straightforward to calculate the Ricci tensor and we find that the metric is Einstein with
$R_{XY}=-\tfrac{3}{2}g_{XY}$.

The $SU(2)$ part of the curvature 2-form is related to the triplet of complex structures via $\vec{R}_{XY}=-\tfrac{1}{4}\vec{J}_{XY}$ (as in e.g. B.70 of \cite{Bergshoeff:2002qk}).
After raising an index via
$\vec{J}_{X}^{\phantom{X}Y}=\vec{J}_{XZ}\,g^{ZY}$ we explicitly find
\begin{align}
(J^{1})^{\phantom{X}Y}_{X}&=\frac{1}{2}\left(
\begin{array}{cccc}
 0 & -4e^{-\varphi } z^1 & 0 & -2e^{-\varphi } \\
 0 & 2e^{\varphi } z^2 & -e^{\varphi } & 0 \\
 0 & 4e^{\varphi } \left(z^2\right)^2+4e^{-\varphi } & -2 e^{\varphi } z^2 & 0 \\
 2e^{\varphi } & -4e^{\varphi } z^1 z^2 & 2e^{\varphi } z^1 & 0 \\
\end{array}
\right)\,,\nn
(J^{2})^{\phantom{X}Y}_{X}&=\frac{1}{2}\left(
\begin{array}{cccc}
 0 & -4e^{-\varphi } z^2 & 2e^{-\varphi } & 0 \\
 0 & -2e^{\varphi } z^1 & 0 & -e^{\varphi } \\
 -2e^{\varphi } & -4e^{\varphi } z^1 z^2 & 0 & -2e^{\varphi } z^2 \\
 0 & 4e^{\varphi } \left(z^1\right)^2+4e^{-\varphi } & 0 & 2 e^{\varphi } z^1 \\
\end{array}
\right)\,,\nn
(J^{3})^{\phantom{X}Y}_{X}&=\frac{1}{2}\left(
\begin{array}{cccc}
 0 & -4e^{-2 \varphi } & 0 & 0 \\
 {e^{2 \varphi }} & 0 & 0 & 0 \\
 2e^{2 \varphi } z^2 & -4z^1 & 0 & -2 \\
 -2 e^{2 \varphi } z^1 & -4z^2 & 2 & 0 \\
\end{array}
\right)\,,
\end{align}
and one can check that $J^i J^j=-\delta^{ij}+\epsilon^{ijk}J^k$.

We are now ready to show that the scalar potential terms in the $SO(2)_D$ truncated theory \eqref{afredfnrwo}
are consistent with $N=2$ supersymmetry. The scalar potential terms in the general $N=2$, $D=5$ gauged supergravity Lagrangian \eqref{bosntwgen} (with no tensor multiplets and no FI terms) are given by
\begin{equation}\label{ntwopotapp}
\mathcal{L}^{pot}_{N=2}=4{g}^2(4\vec{P}\cdot\vec{P}-2\vec{P}^{x}\cdot \vec{P}_{x}-2W_{x}W^x-2\mathcal{N}_{A}\mathcal{N}^{A})\,.
\end{equation}
Lets discuss each of these terms. The first two terms involve the moment maps for the 
Killing vectors 
$k^X_{I}$ defined via
\begin{equation}
\vec{P}_{I}=\frac{1}{2}\vec{J}_{X}^{\phantom{X}Y}\nabla_Y k^X_{I}\,.
\end{equation}
The terms appearing in the scalar potential are then determined by
\begin{equation}
\vec{P}\equiv  \tfrac{1}{2}h^I \vec{P}_{I}\,,\qquad  \vec{P}_{x}\equiv  \tfrac{1}{2}h^I_{x}\vec{P}_{I}\,,
\end{equation}
where
\begin{align}
h_x^I=-\sqrt{3}\partial_x h^I\,,
\end{align}
and indices are raised and lowered using the metrics $g_{xy}$ and $a_{IJ}$
given in \eqref{aijmet},\eqref{gxymet}.
For the explicit Killing vectors of the metric \eqref{quatmetap} given by
\begin{equation}\label{kvecsfinap}
k_0=z^2\partial_1-z^1\partial_2\,,\qquad
k_1=l\partial_\xi+z^1\partial_2-z^2\partial_1\,,
\qquad
k_2=\partial_\xi\,,
\end{equation}
we find
\begin{align}
\vec{P}_0=&\left(-2e^{\varphi } z^1,-2e^{\varphi } z^2,-1+e^{2 \varphi} z^az^a\right)\,,\nn
\vec{P}_1=&\left(2e^{\varphi} z^1,2e^{\varphi} z^2, 1+\tfrac{1}{2} e^{2 \varphi} \left(l-2 z^az^a\right)\right)\,,\nn
\vec{P}_2=&\left(0,0,\tfrac{1}{2}e^{2\varphi}\right)\,.
\end{align}

We next note that without tensor multiplets we have
\begin{equation}
W^x\equiv -\tfrac{3}{4}\bar f_{IJ}{}^Kh^I h^J h_K^x\,,
\end{equation}
where $\bar f_{IJ}{}^K$ are the structure constants for the gauging. 
For our gauging we have $\bar f_{IJ}{}^K=0$ and hence $W^x=0$. The final terms
in the scalar potential are given by
\begin{equation}
\mathcal{N}_{A}\mathcal{N}^{A}\equiv \tfrac{3}{16}h^I k^X_{I}g_{XY}h^Jk^Y_{J}\,.
\end{equation}
After explicitly evaluating the terms in \eqref{ntwopotapp} using the ingredients in this appendix as well as those in
section \ref{so2dsec}, we precisely recover the scalar potential terms in \eqref{afredfnrwo}.


\begin{thebibliography}{10}

\bibitem{Gauntlett:2007ma}
J.~P. Gauntlett and O.~Varela, ``{Consistent Kaluza-Klein Reductions for
  General Supersymmetric AdS Solutions},''
  \href{http://dx.doi.org/10.1103/PhysRevD.76.126007}{{\em Phys. Rev.}
  {\bfseries D76} (2007) 126007},
\href{http://arxiv.org/abs/0707.2315}{{\ttfamily arXiv:0707.2315 [hep-th]}}.

\bibitem{Nastase:1999cb}
H.~Nastase, D.~Vaman, and P.~van Nieuwenhuizen, ``{Consistent nonlinear K K
  reduction of 11-d supergravity on AdS(7) x S(4) and selfduality in odd
  dimensions},'' \href{http://dx.doi.org/10.1016/S0370-2693(99)01266-6}{{\em
  Phys. Lett.} {\bfseries B469} (1999) 96--102},
\href{http://arxiv.org/abs/hep-th/9905075}{{\ttfamily arXiv:hep-th/9905075
  [hep-th]}}.

\bibitem{Nastase:1999kf}
H.~Nastase, D.~Vaman, and P.~van Nieuwenhuizen, ``{Consistency of the AdS(7) x
  S(4) reduction and the origin of selfduality in odd dimensions},''
  \href{http://dx.doi.org/10.1016/S0550-3213(00)00193-0}{{\em Nucl. Phys.}
  {\bfseries B581} (2000) 179--239},
\href{http://arxiv.org/abs/hep-th/9911238}{{\ttfamily arXiv:hep-th/9911238
  [hep-th]}}.

\bibitem{deWit:1986oxb}
B.~de~Wit and H.~Nicolai, ``{The Consistency of the S**7 Truncation in D=11
  Supergravity},''
\href{http://dx.doi.org/10.1016/0550-3213(87)90253-7}{{\em Nucl. Phys.}
  {\bfseries B281} (1987) 211--240}.

\bibitem{Cvetic:2000nc}
M.~Cvetic, H.~Lu, C.~Pope, A.~Sadrzadeh, and T.~A. Tran, ``{Consistent SO(6)
  reduction of type IIB supergravity on $S^5$},''
  \href{http://dx.doi.org/10.1016/S0550-3213(00)00372-2}{{\em Nucl.Phys.}
  {\bfseries B586} (2000) 275--286},
  \href{http://arxiv.org/abs/hep-th/0003103}{{\ttfamily arXiv:hep-th/0003103
  [hep-th]}}.

\bibitem{Lee:2014mla}
K.~Lee, C.~Strickland-Constable, and D.~Waldram, ``{Spheres, generalised
  parallelisability and consistent truncations},''
  \href{http://dx.doi.org/10.1002/prop.201700048}{{\em Fortsch. Phys.}
  {\bfseries 65} no.~10-11, (2017) 1700048},
\href{http://arxiv.org/abs/1401.3360}{{\ttfamily arXiv:1401.3360 [hep-th]}}.

\bibitem{Baguet:2015sma}
A.~Baguet, O.~Hohm, and H.~Samtleben, ``{Consistent Type IIB Reductions to
  Maximal 5D Supergravity},''
  \href{http://dx.doi.org/10.1103/PhysRevD.92.065004}{{\em Phys. Rev.}
  {\bfseries D92} no.~6, (2015) 065004},
\href{http://arxiv.org/abs/1506.01385}{{\ttfamily arXiv:1506.01385 [hep-th]}}.

\bibitem{Maldacena:2000mw}
J.~M. Maldacena and C.~Nunez, ``{Supergravity description of field theories on
  curved manifolds and a no go theorem},''
  \href{http://dx.doi.org/10.1142/S0217751X01003937}{{\em Int.J.Mod.Phys.}
  {\bfseries A16} (2001) 822--855},
\href{http://arxiv.org/abs/hep-th/0007018}{{\ttfamily arXiv:hep-th/0007018
  [hep-th]}}.

\bibitem{Gauntlett:2007sm}
J.~P. Gauntlett and O.~Varela, ``{D=5 $SU(2)$xU(1) Gauged Supergravity from
  D=11 Supergravity},''
  \href{http://dx.doi.org/10.1088/1126-6708/2008/02/083}{{\em JHEP} {\bfseries
  02} (2008) 083},
\href{http://arxiv.org/abs/0712.3560}{{\ttfamily arXiv:0712.3560 [hep-th]}}.

\bibitem{Bobev:2018sgr}
N.~Bobev, D.~Cassani, and H.~Triendl, ``{Holographic RG Flows for
  Four-dimensional $\mathcal{N}=2$ SCFTs},''
  \href{http://dx.doi.org/10.1007/JHEP06(2018)086}{{\em JHEP} {\bfseries 06}
  (2018) 086},
\href{http://arxiv.org/abs/1804.03276}{{\ttfamily arXiv:1804.03276 [hep-th]}}.

\bibitem{Gauntlett:2002rv}
J.~P. Gauntlett, N.~Kim, S.~Pakis, and D.~Waldram, ``{M theory solutions with
  AdS factors},'' \href{http://dx.doi.org/10.1088/0264-9381/19/15/305}{{\em
  Class. Quant. Grav.} {\bfseries 19} (2002) 3927--3946},
\href{http://arxiv.org/abs/hep-th/0202184}{{\ttfamily arXiv:hep-th/0202184
  [hep-th]}}.

\bibitem{Szepietowski:2012tb}
P.~Szepietowski, ``{Comments on a-maximization from gauged supergravity},''
  \href{http://dx.doi.org/10.1007/JHEP12(2012)018}{{\em JHEP} {\bfseries 12}
  (2012) 018},
\href{http://arxiv.org/abs/1209.3025}{{\ttfamily arXiv:1209.3025 [hep-th]}}.

\bibitem{Pernici:1984xx}
M.~Pernici, K.~Pilch, and P.~van Nieuwenhuizen, ``{Gauged Maximally Extended
  Supergravity in Seven-dimensions},''
\href{http://dx.doi.org/10.1016/0370-2693(84)90813-X}{{\em Phys. Lett.}
  {\bfseries 143B} (1984) 103--107}.

\bibitem{Cvetic:2000ah}
M.~Cvetic, H.~Lu, C.~N. Pope, A.~Sadrzadeh, and T.~A. Tran, ``{S**3 and S**4
  reductions of type IIA supergravity},''
  \href{http://dx.doi.org/10.1016/S0550-3213(00)00466-1}{{\em Nucl. Phys.}
  {\bfseries B590} (2000) 233--251},
\href{http://arxiv.org/abs/hep-th/0005137}{{\ttfamily arXiv:hep-th/0005137
  [hep-th]}}.

\bibitem{Schon:2006kz}
J.~Schon and M.~Weidner, ``{Gauged N=4 supergravities},''
  \href{http://dx.doi.org/10.1088/1126-6708/2006/05/034}{{\em JHEP} {\bfseries
  05} (2006) 034},
\href{http://arxiv.org/abs/hep-th/0602024}{{\ttfamily arXiv:hep-th/0602024
  [hep-th]}}.

\bibitem{DallAgata:2001wgl}
G.~Dall'Agata, C.~Herrmann, and M.~Zagermann, ``{General matter coupled N=4
  gauged supergravity in five-dimensions},''
  \href{http://dx.doi.org/10.1016/S0550-3213(01)00367-4}{{\em Nucl. Phys.}
  {\bfseries B612} (2001) 123--150},
\href{http://arxiv.org/abs/hep-th/0103106}{{\ttfamily arXiv:hep-th/0103106
  [hep-th]}}.

\bibitem{Awada:1985ep}
M.~Awada and P.~K. Townsend, ``{$N=4$ Maxwell-einstein Supergravity in
  Five-dimensions and Its SU(2) Gauging},''
\href{http://dx.doi.org/10.1016/0550-3213(85)90156-7}{{\em Nucl. Phys.}
  {\bfseries B255} (1985) 617--632}.

\bibitem{Lu:1998xt}
H.~Lu, C.~N. Pope, and K.~S. Stelle, ``{M theory / heterotic duality: A
  Kaluza-Klein perspective},''
  \href{http://dx.doi.org/10.1016/S0550-3213(99)00086-3}{{\em Nucl. Phys.}
  {\bfseries B548} (1999) 87--138},
\href{http://arxiv.org/abs/hep-th/9810159}{{\ttfamily arXiv:hep-th/9810159
  [hep-th]}}.

\bibitem{Romans:1985ps}
L.~J. Romans, ``{Gauged N=4 supergravities in five-dimensions and their
  magnetovac backgrounds},''
\href{http://dx.doi.org/10.1016/0550-3213(86)90398-6}{{\em Nucl. Phys.}
  {\bfseries B267} (1986) 433}.

\bibitem{Bergshoeff:2004kh}
E.~Bergshoeff, S.~Cucu, T.~de~Wit, J.~Gheerardyn, S.~Vandoren, and
  A.~Van~Proeyen, ``{N = 2 supergravity in five-dimensions revisited},''
  \href{http://dx.doi.org/10.1088/0264-9381/23/23/C01,
  10.1088/0264-9381/21/12/013}{{\em Class. Quant. Grav.} {\bfseries 21} (2004)
  3015--3042}, \href{http://arxiv.org/abs/hep-th/0403045}{{\ttfamily
  arXiv:hep-th/0403045 [hep-th]}}.
[Class. Quant. Grav.23,7149(2006)].

\bibitem{Gunaydin:1984ak}
M.~Gunaydin, G.~Sierra, and P.~K. Townsend, ``{Gauging the d = 5
  Maxwell-Einstein Supergravity Theories: More on Jordan Algebras},''
\href{http://dx.doi.org/10.1016/0550-3213(85)90547-4}{{\em Nucl. Phys.}
  {\bfseries B253} (1985) 573}.

\bibitem{Gunaydin:1999zx}
M.~Gunaydin and M.~Zagermann, ``{The Gauging of five-dimensional, N=2
  Maxwell-Einstein supergravity theories coupled to tensor multiplets},''
  \href{http://dx.doi.org/10.1016/S0550-3213(99)00801-9}{{\em Nucl. Phys.}
  {\bfseries B572} (2000) 131--150},
\href{http://arxiv.org/abs/hep-th/9912027}{{\ttfamily arXiv:hep-th/9912027
  [hep-th]}}.

\bibitem{Ceresole:2000jd}
A.~Ceresole and G.~Dall'Agata, ``{General matter coupled N=2, D = 5 gauged
  supergravity},'' \href{http://dx.doi.org/10.1016/S0550-3213(00)00339-4}{{\em
  Nucl. Phys.} {\bfseries B585} (2000) 143--170},
\href{http://arxiv.org/abs/hep-th/0004111}{{\ttfamily arXiv:hep-th/0004111
  [hep-th]}}.

\bibitem{Gunaydin:2000ph}
M.~Gunaydin and M.~Zagermann, ``{Gauging the full R symmetry group in
  five-dimensional, N=2 Yang-Mills Einstein tensor supergravity},''
  \href{http://dx.doi.org/10.1103/PhysRevD.63.064023}{{\em Phys. Rev.}
  {\bfseries D63} (2001) 064023},
\href{http://arxiv.org/abs/hep-th/0004117}{{\ttfamily arXiv:hep-th/0004117
  [hep-th]}}.

\bibitem{Cordova:2016emh}
C.~Cordova, T.~T. Dumitrescu, and K.~Intriligator, ``{Multiplets of
  Superconformal Symmetry in Diverse Dimensions},''
  \href{http://dx.doi.org/10.1007/JHEP03(2019)163}{{\em JHEP} {\bfseries 03}
  (2019) 163},
\href{http://arxiv.org/abs/1612.00809}{{\ttfamily arXiv:1612.00809 [hep-th]}}.

\bibitem{Louis:2015dca}
J.~Louis, H.~Triendl, and M.~Zagermann, ``{$ \mathcal{N}=4 $ supersymmetric
  AdS$_{5}$ vacua and their moduli spaces},''
  \href{http://dx.doi.org/10.1007/JHEP10(2015)083}{{\em JHEP} {\bfseries 10}
  (2015) 083},
\href{http://arxiv.org/abs/1507.01623}{{\ttfamily arXiv:1507.01623 [hep-th]}}.

\bibitem{Nieder:2000kc}
H.~Nieder and Y.~Oz, ``{Supergravity and D-branes wrapping special Lagrangian
  cycles},'' \href{http://dx.doi.org/10.1088/1126-6708/2001/03/008}{{\em JHEP}
  {\bfseries 03} (2001) 008},
\href{http://arxiv.org/abs/hep-th/0011288}{{\ttfamily arXiv:hep-th/0011288
  [hep-th]}}.

\bibitem{Gauntlett:2001jj}
J.~P. Gauntlett and N.~Kim, ``{M five-branes wrapped on supersymmetric cycles.
  2.},'' \href{http://dx.doi.org/10.1103/PhysRevD.65.086003}{{\em Phys. Rev.}
  {\bfseries D65} (2002) 086003},
\href{http://arxiv.org/abs/hep-th/0109039}{{\ttfamily arXiv:hep-th/0109039
  [hep-th]}}.

\bibitem{Caldarelli:1998hg}
M.~M. Caldarelli and D.~Klemm, ``{Supersymmetry of Anti-de Sitter black
  holes},'' \href{http://dx.doi.org/10.1016/S0550-3213(98)00846-3}{{\em Nucl.
  Phys.} {\bfseries B545} (1999) 434--460},
\href{http://arxiv.org/abs/hep-th/9808097}{{\ttfamily arXiv:hep-th/9808097
  [hep-th]}}.

\bibitem{Gauntlett:2000ng}
J.~P. Gauntlett, N.~Kim, and D.~Waldram, ``{M-fivebranes wrapped on
  supersymmetric cycles},''
  \href{http://dx.doi.org/10.1103/PhysRevD.63.126001}{{\em Phys. Rev.}
  {\bfseries D63} (2001) 126001},
\href{http://arxiv.org/abs/hep-th/0012195}{{\ttfamily arXiv:hep-th/0012195}}.

\bibitem{privcom}
N.~Bobev {\em {, Private communication}} .

\bibitem{Benini:2012cz}
F.~Benini and N.~Bobev, ``{Exact two-dimensional superconformal R-symmetry and
  c-extremization},''
  \href{http://dx.doi.org/10.1103/PhysRevLett.110.061601}{{\em Phys. Rev.
  Lett.} {\bfseries 110} no.~6, (2013) 061601},
\href{http://arxiv.org/abs/1211.4030}{{\ttfamily arXiv:1211.4030 [hep-th]}}.

\bibitem{Benini:2013cda}
F.~Benini and N.~Bobev, ``{Two-dimensional SCFTs from wrapped branes and
  c-extremization},'' \href{http://dx.doi.org/10.1007/JHEP06(2013)005}{{\em
  JHEP} {\bfseries 06} (2013) 005},
\href{http://arxiv.org/abs/1302.4451}{{\ttfamily arXiv:1302.4451 [hep-th]}}.

\bibitem{Comsa:2019rcz}
I.~M. Comsa, M.~Firsching, and T.~Fischbacher, ``{SO(8) Supergravity and the
  Magic of Machine Learning},''
  \href{http://dx.doi.org/10.1007/JHEP08(2019)057}{{\em JHEP} {\bfseries 08}
  (2019) 057},
\href{http://arxiv.org/abs/1906.00207}{{\ttfamily arXiv:1906.00207 [hep-th]}}.

\bibitem{Donos:2010ax}
A.~Donos, J.~P. Gauntlett, N.~Kim, and O.~Varela, ``{Wrapped M5-branes,
  consistent truncations and AdS/CMT},''
  \href{http://dx.doi.org/10.1007/JHEP12(2010)003}{{\em JHEP} {\bfseries 12}
  (2010) 003},
\href{http://arxiv.org/abs/1009.3805}{{\ttfamily arXiv:1009.3805 [hep-th]}}.

\bibitem{Hohm:2014qga}
O.~Hohm and H.~Samtleben, ``{Consistent Kaluza-Klein Truncations via
  Exceptional Field Theory},''
  \href{http://dx.doi.org/10.1007/JHEP01(2015)131}{{\em JHEP} {\bfseries 01}
  (2015) 131},
\href{http://arxiv.org/abs/1410.8145}{{\ttfamily arXiv:1410.8145 [hep-th]}}.

\bibitem{Malek:2017njj}
E.~Malek, ``{Half-Maximal Supersymmetry from Exceptional Field Theory},''
  \href{http://dx.doi.org/10.1002/prop.201700061}{{\em Fortsch. Phys.}
  {\bfseries 65} no.~10-11, (2017) 1700061},
\href{http://arxiv.org/abs/1707.00714}{{\ttfamily arXiv:1707.00714 [hep-th]}}.

\bibitem{Cassani:2019vcl}
D.~Cassani, G.~Josse, M.~Petrini, and D.~Waldram, ``{Systematics of consistent
  truncations from generalised geometry},''
\href{http://arxiv.org/abs/1907.06730}{{\ttfamily arXiv:1907.06730 [hep-th]}}.

\bibitem{Strominger:1997eb}
A.~Strominger, ``{Loop corrections to the universal hypermultiplet},''
  \href{http://dx.doi.org/10.1016/S0370-2693(98)00015-X}{{\em Phys. Lett.}
  {\bfseries B421} (1998) 139--148},
\href{http://arxiv.org/abs/hep-th/9706195}{{\ttfamily arXiv:hep-th/9706195
  [hep-th]}}.

\bibitem{Lukas:1998tt}
A.~Lukas, B.~A. Ovrut, K.~S. Stelle, and D.~Waldram, ``{Heterotic M theory in
  five-dimensions},''
  \href{http://dx.doi.org/10.1016/S0550-3213(99)00196-0}{{\em Nucl. Phys.}
  {\bfseries B552} (1999) 246--290},
\href{http://arxiv.org/abs/hep-th/9806051}{{\ttfamily arXiv:hep-th/9806051
  [hep-th]}}.

\bibitem{Bergshoeff:2002qk}
E.~Bergshoeff, S.~Cucu, T.~De~Wit, J.~Gheerardyn, R.~Halbersma, S.~Vandoren,
  and A.~Van~Proeyen, ``{Superconformal N=2, D = 5 matter with and without
  actions},'' \href{http://dx.doi.org/10.1088/1126-6708/2002/10/045}{{\em JHEP}
  {\bfseries 10} (2002) 045},
\href{http://arxiv.org/abs/hep-th/0205230}{{\ttfamily arXiv:hep-th/0205230
  [hep-th]}}.

\end{thebibliography}
\providecommand{\href}[2]{#2}\begingroup\raggedright\endgroup

\end{document}